\documentclass[a4paper]{article}
\usepackage[T2A,T1]{fontenc}
\usepackage[english]{babel}
\usepackage{amsmath}
\usepackage{amssymb}
\usepackage{caption}
\usepackage{hyperref}
\hypersetup{colorlinks=true, linkcolor=blue, citecolor=blue, filecolor=blue, urlcolor=blue}
\usepackage{array}
\usepackage{hhline}
\usepackage{graphicx}

\title{Review of scientific topics for Millimetron space observatory}
\date{}

\begin{document}
\maketitle

\textit{N.S.~Kardashev$^1$, I.D.~Novikov$^{1,2}$, V.N.~Lukash$^1$, S.V.~Pilipenko$^1$, E.V.~Mikheeva$^1$, D.V.~Bisikalo$^3$, D.S.~Wiebe$^3$, A.G.~Doroshkevich$^1$, A.V.~Zasov$^4$, I.I.~Zinchenko$^{5,6}$, P.B.~Ivanov$^1$, V.I.~Kostenko$^1$, T.I.~Larchenkova$^1$, S.F.~Likhachev$^1$, I.F.~Malov$^7$, V.M.~Malofeev$^7$, A.S.~Pozanenko$^8$, A.V.~Smirnov$^1$, A.M.~Sobolev$^9$, A.M.~Cherepashchuk$^4$, Yu.A.~Shchekinov$^{10}$}
\vskip12pt
\noindent$^1$Lebedev Physical Institute, AstroSpace Center, Moscow, Russia\\
$^2$The Niels Bohr Institute, Copenhagen, Denmark\\
$^3$Institute of Astronomy of the Russian Academy of Sciences (INASAN), Moscow, Russia\\
$^4$Lomonosov Moscow State University Sternberg Astronomical Institute, Moscow, Russia\\
$^5$Institute of Applied Physics, Nizhnii Novgorod, Russia\\
$^6$N.I. Lobachevskii Nizhnii Novgorod State University, Nizhnii Novgorod, Russia\\
$^7$Lebedev Physical Institute, Pushchino RadioAstronomical Observatory, Pushchino, Russia\\
$^8$Space Research Institute, Moscow, Russia\\
$^9$Ural Federal University, Institute of Natural Sciences, Astronomical Observatory, Ekaterinburg, Russia\\
$^{10}$South Federal University, Rostov-on-Don, Russia\\

\begin{center}
\textit{Abstract}
\end{center}

This paper describes outstanding issues in astrophysics and cosmology that can be solved by astronomical observations in a broad spectral range from far infrared to millimeter wavelengths.  The discussed problems related to the formation of stars and planets, galaxies and the interstellar medium, studies of black holes and the development of the cosmological model can be addressed by the planned space observatory Millimetron (the ``Spectr-M'' project) equipped with a cooled 10-m mirror. Millimetron can operate both as a single-dish telescope and as a part of a space-ground interferometer with very long baseline.

\setcounter{tocdepth}{10}
\renewcommand\contentsname{Contents}
\tableofcontents
\clearpage

\section{Introduction}

The Millimetron space observatory is aimed at astronomical observations, which address a broad range of objects in the Universe in the  20 $\mu$m to 20 mm wavelength range. Since the beginning of 1990s, there has been a growing astrophysical and cosmological interest to objects in this range.  This range is very important to different types of astronomical observations, including the continuum and spectral line studies, polarimetry and variations of different parameters.

The millimeter (1 mm $<\lambda<$ 1 cm), sub-millimeter (0.1 $<\lambda<$ 1 mm) and far infrared (FIR) (50 $<\lambda<$ 300 $\mu$m) bands are unique for astronomical observations for the following reasons:

\begin{itemize}
\item the cosmic microwave background (CMB) peaks at 1 mm wavelength. This is the only electromagnetic radiation that survived since the earliest stages of the Universe (the Big Bang) and fills the space almost homogeneously. A detailed study of the space structure, spectrum and polarization of CMB will help in solving fundamental issues in astrophysics and cosmology, including the development of the standard cosmological model, the origin and evolution of the first objects in the Universe, determination of dark matter and dark energy parameters, etc.;

\item in FIR, there is a maximum emission from the coldest objects in the Universe, including gas and dust clouds in our and other galaxies, asteroids, comets and planets. Therefore, observations in this range allow exploring the interstellar medium evolution in the process of the gravitational contraction leading to the formation of stars and planetary systems, and ultimately to the appearance of life and civilizations;

\item the sky background emission reaches minimum near 300 $\mu$m. This background is a sum of contributions from different sources along the line of sight and CMB (Fig. 1a). Therefore, observations in this range will enable probing very faint astronomical objects --- galaxies, stars, black holes, exoplanets, etc., --- with the highest sensitivity;

\item in the submillimeter  and FIR ranges there is a lot of atomic and molecular spectral lines, allowing the determination of the chemical composition and physical properties of gas in different objects ranging from protoplanetary disks to galaxies at different epochs;

\item submillimeter observations can significantly increase the angular resolution using Very Long Baseline Interferometry (VLBI), which is necessary to study the most compact objects, such as  surroundings of black holes, some pulsars and gamma-ray bursts;

\item the medium surrounding many interesting astronomical objects is mostly transparent in these wavelength ranges compared to nearby spectral bands, both at short wavelengths (due to the interstellar dust absorption) and at long wavelengths (due to synchrotron self-absorption, thermal plasma absorption and scattering on plasma inhomogeneities).
\end{itemize}

One of the largest single ground-based telescopes specifically designed for submillimeter observations is the James Clerk Maxwell Telescope, JCMT (\url{http://www.jach.hawaii.edu/JCMT/}) with an aperture 15 m located near Mauna Kea (Hawaii) at an altitude of more than 4000 m from the sea level. Heterodyne detectors of JCMT cover all windows of the Earth atmosphere transparency in the 200-700 GHz frequency band, and the bolometer array detector consisting of 10240 elements operates at two wavelengths: 450 and 850 $\mu$m. JCMT is used to study the Solar System, interstellar gas and dust, as well as distant galaxies.  Using JCMT, submillimeter galaxies were discovered in which submillimeter emission dominates over the optical emission.

Atacama Large Millimeter/submillimeter Array (ALMA) is the most   prospective ground-based submillimeter instrument (\url{http://www.almaobservatory.org}), which is now at the commissioning stage. It represents a compact interferometer with a base up to 16 km consisting of 66 antennas, of which 54 have diameter of 12 meters and 12 of 7 meters. The ALMA observatory is located at the altitude of 5000 m above the sea level in the Atakama desert in Chile and is capable of conducting astronomical observations in all transparency windows from 10 mm to 0.3 mm. The fully operational ALMA observatory will provide an unprecedentedly high sensitivity of up to 50 $\mu$Jy in the continuum with an angular resolution of less than 0.1 arcsec, which allows a detailed mapping of protoplanetary disks and studies of the morphology of distant galaxies. A small field of view is a shortcoming of ALMA, which requires long observational time to carry out large surveys of point-like sources, studies of extended star formation regions in the Galaxy and mapping of large sky areas.

\begin{figure}
\begin{center}
\includegraphics[width=7.5cm]{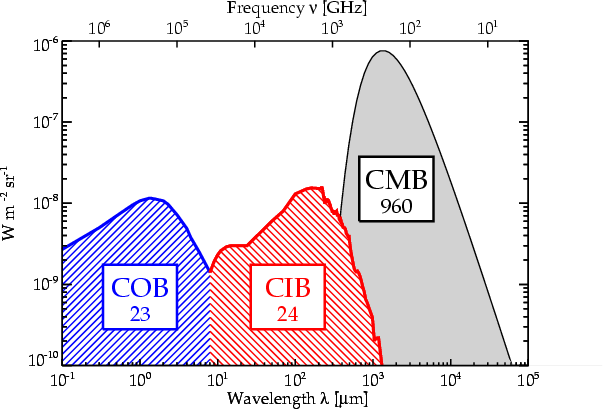}
\includegraphics[width=7.5cm]{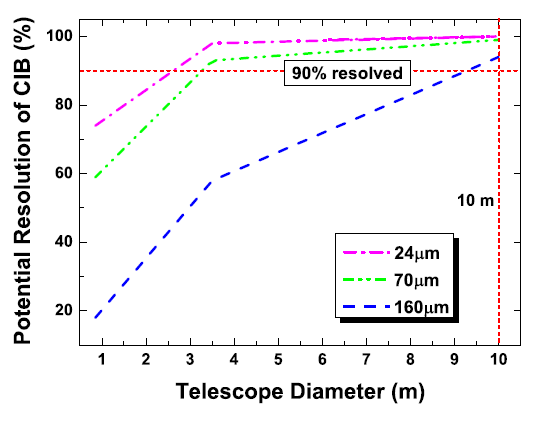}
\end{center}
\caption{(a) Extragalactic background spectrum. COB --- cosmic optical background, CIB --- cosmic infrared background, CMB --- cosmic microwave background. Numbers show the total intensity of the background components in units nW m$^2$ sr$^{-1}$ [1]. (b) The possible fraction of CIB resolved into individual sources as a function of the telescope diameter for several wavelengths shown in the Figure [2].}
\end{figure}

As noted above, ground-based astronomical observations at wavelengths $\lambda \lesssim 300$ $\mu$m are problematic, since they are significantly limited by  properties of the terrestrial atmosphere: its proper emission and absorption by water vapour, oxygen, carbon dioxide and ozone. Therefore, observations at wavelengths shorter than 300 $\mu$m should be carried out at altitudes at least as high as 10-30 km, where partial water vapour pressure strongly decreases with altitude, and the associated absorption almost vanishes. This property has already been exploited by telescopes installed on airplanes (SOFIA (Stratospheric Observatory for Infrared Astronomy) [3]) or stratospheric balloons (BOOMERANG (Balloon Observations Of Millimeter Extragalactic Radiation and Geophysics) [4], TELIS (TEraHertz and submillimeter  Limb Sounder) [5], Olimpo, etc.). A space telescope has clear advantages, since it is free from the negative effect of the Earth atmosphere and can be cooled down to low temperatures, thus strongly increasing its sensitivity.

The Herschel space observatory launched in 2009 and operated until the middle of 2013 is the most perfect and close predecessor of Millimetron observatory. The Herschel (\url{http://sci.esa.int/Herschel/} [6]) consists of a 3.5-m diameter telescope with passive cooling down to about 70 K. The observatory carried out observations in the wavelength range from 55 to 672 $\mu$m. The receivers of the Herschel included sensitive array photometers, an array spectrometer with moderate resolution and a high-resolution heterodyne spectrometer. The main scientific achievements of the Herschel observatory include observations of star-forming and dust regions in our and external galaxies, studies of submillimeter galaxies  and Solar system bodies.

\begin{table}
\caption{Main parameters of the Millimetron observatory}
\hspace{-1cm}
\begin{tabular}{|m{3cm}|m{5.8cm}|m{7.2cm}|}
\hline
{\bfseries Mode} &
Single dish &
Space-Earth interferometer\\\hline
{\bfseries Possibilities } &
\begin{itemize}
\item Low spectral resolution or photometry (R=$\lambda $/$\Delta \lambda \sim$3).
\item Medium spectral resolution ($R\sim10^{3}$).
\item High spectral resolution ($R{\geq}10^{6}$).
\item Polarimetry.
\end{itemize}
 &
\begin{itemize}
\item Estimation of source angular size
\item One dimensional source cross-section
\item Maps with a priori source model adopted
\end{itemize}
\\\hline
{\bfseries Working wavelength range} &
20 $\mu$m -- 3 mm (for R$\sim3$ and R$\sim 10^{3}$), 60 $\mu$m -- 0.6 mm (for R${\geq}10^{6}$) &
0.3 -- 17 mm

\\\hline
{\bfseries Angular resolution} &
6 arcsec  (at $\lambda $ = 300 $\mu$m) &
for baseline 1.5 million km:

$\sim${}2 $\mu$arcsec, ($\lambda $ = 13.5 mm),

$\sim${}50 narcsec,($\lambda $ = 0.345 mm)

\\\hline
{\bfseries Sensitivity }

 &
Noise RMS $\sigma$ (for ${\lambda}$ = 300 $\mu$m, integration time t= 3600 s, effective area A=50 m$^2$, detector sensitivity NEP$^*$ ${\leq 10^{-19}}$
W Hz$^{-1/2}$):

\begin{itemize}
\item 20 nJy (for R$\sim${}3)
\item 4 $\mu$Jy (or $4\cdot 10^{-23}$ W m$^-2$ (for R$\sim 10^3$)
\end{itemize}
 &
$\sigma $ for bandwidth 4GHz (2 polarizations, 2 bit quantization)
\begin{flushleft}
\begin{tabular}{|m{2.0809999cm}|m{2.0809999cm}|m{1.8cm}|}
\hline
Frequency, GHz &
Coherent integration time [9], s &
Sensitivity, mJy\\\hline
22 &
500 &
0.2\\\hline
43 &
300 &
0.3\\\hline
100 &
100 &
1.0\\\hline
240 &
70 &
1.9\\\hline
640 &
10 &
20.0\\\hline
870 &
5 &
40.0\\\hline
\end{tabular}
\end{flushleft}
\\\hline

\multicolumn{3}{|l|}{$^*$ NEP -- Noise Equivalent Power}
\\\hline
\end{tabular}
\end{table}

Millimetron space observatory represents a new step in the development of space FIR missions due to its unique characteristics: high angular resolution and unprecedentedly high sensitivity in a broad wavelength range from far infrared to millimeters. This leap forward may solve many fundamental issues of astrophysics and cosmology. The unique breakthrough scientific tasks are crucial in determining technical characteristics of the Millimetron observatory.

Millimetron  will be launched in an orbit in the vicinity of the Lagrangian point L2 at a distance of 1.5 million km from Earth behind the Moon orbit, with the most favorable external conditions for the telescope cooling.  To provide an unprecedentedly high sensitivity, a deep cooling of the telescope mirrors down to temperatures as low as 10 K is required. Such a regime can be realized only by joint operation of two cooling systems: passive and active. The former utilizes the solar shields, while the second one uses close-cycle cryogenic refrigerators. Being in the vicinity of the Lagrangian L2 point, Millimetron forms together with a ground-based telescope or a system of telescopes -- an interferometer with maximum projection of the base on the plane perpendicular to the line of sight of a studied source of more than 1.5 million km. Such a unique instrument opens new horizons in solving astrophysical problems by enabling measurements with record high angular resolution.

The Spectr-M project started in the 1990s. A description of different concepts of the construction of the Millimetron observatory since the beginning of the project and a list of its main science tasks can be found in papers [2, 7, 8]. The main parameters of the Millimetron are listed in Table 1.

It should be noted that Table 1 presents the desirable parameters, which can be realized at the present level of technology. The characteristics of the real observatory will be recommended according to a careful compromise between the desirable parameters, the priority of scientific tasks and the project cost.

\begin{table}
\caption{Estimated parameters of the Millimetron observatory in the single dish mode and the main characteristics of other telescopes and observatories}
\begin{center}
\begin{tabular}{|l|c|c|c|c|}
\hline
 &
{\bfseries Herschel} &
{\bfseries ALMA} &
{\bfseries SPICA} &
{\bfseries Millimetron}\\\hline
Range, $\mu$m &
50-670 &
315-9680 &
5-210 &
20-3000\\\hline
resolution, arcsec &
3.5-40 &
0.01-5 &
0.3-14 &
5-60\\\hline
field of view &
up to 4$^\prime\times8^\prime$ &
up to 25$^{\prime\prime}$ &
$5^\prime\times5^\prime$ &
$6^\prime\times6^\prime$\\\hline
\multicolumn{5}{|c|}{\centering{sensitivity $\sigma $ for the integration time t = 3600
s*}}\\\hline
photometry &
1 mJy &
{\textgreater}10 $\mu$Jy &
4 $\mu$Jy &
20 nJy\\\hline
spectroscopy (R$\sim${}1000) &
20 mJy &
60 $\mu$Jy &
200 $\mu$Jy &
4 $\mu$Jy\\\hline
spectroscopy (R${\geq}10^6$) &
2 Jy &
50 mJy &
{}- &
200 mJy\\\hline
\multicolumn{5}{|l|}{* sensitivity depends on wavelength, the sensitivity for $\lambda\sim200-300$ $\mu$m is given.}\\\hline
\end{tabular}
\end{center}
\end{table}

The comparison of the expected parameters of the Millimetron observatory with those of existing and planned instruments for observations in nearby or similar wavelength ranges  (Table 2), such as the ground-based ALMA observatory, the Herschel and SPICA (Space Infrared Telescope for Cosmology and Astrophysics) [10] space telescopes, suggests the list of the highest-priority scientific tasks and allows the formulation of the priority of observations.

The expected sensitivity of Millimetron is at least two orders of magnitude higher than that of the Herschel telescope. Currently the best angular resolution in the 20-300 $\mu$m range is much worse than in other ranges (radio and near-IR). This is due to the fact that the 20-300$\mu$m  range is virtually inaccessible for ground-based observations, and all space telescopes launched so far had small diameters of order of 1 m. Presently, the Herschel space telescope has the largest mirror (diameter of 3.5 m) optimized for observations in far infrared. For deeper studies of different astronomical objects a better angular resolution is needed, and the Millimetron observatory is planned to be the next step by providing three times higher angular resolution in the far-IR range.

The angular resolution of the telescope is also related to the well-known astronomical confusion problem: at a low angular resolution distant sources merge into a homogeneous background, which hampers measurements of individual fainter objects. In the far-IR range this background is usually called Cosmic Infrared Background (CIB). CIB is thought to be mainly due to emission of distant galaxies. Preliminary estimates show that Millimetron with the aperture 10-m will be capable of resolving more than 90\%  of CIB into individual sources --- distant galaxies (Fig. 1b).

At wavelengths longer than 300 $\mu$m ALMA has a better angular resolution than Millimetron, and the ALMA sensitivity at longer wavelengths ($\lambda >$ 1 mm) is also higher than that of Millimetron due to a huge collective area. However, at wavelengths shorter than 300 $\mu$m, which are inaccessible for ALMA, Millimetron will have no competitors in sensitivity. A wide field of view of Millimetron is another advantage. This field of view is provided by several thousands of detectors and enables Millimetron to map large sky areas. A grating spectrometer will offer broad-band spectral measurements, facilitating  measurement of redshifts of distant galaxies, which is a difficult task for ALMA.

Clearly, some tasks observations by ALMA and Millimetron can and should complement each other. In addition, another ambitious project --- James Webb Space Telescope (JWST) (http://www.jwst.nasa.gov) operating at shorter wavelengths, can be an interesting completion. For example, in studies of high-redshift galaxies, Millimetron can analyze a relatively cold molecular and atomic gas, JWST can study properties of a hotter atomic gas, and ALMA will provide a detailed imaging of these galaxies.

In the interferometer mode, the Millimetron observatory can cooperate with the presently developing Event Horizon Telescope (EHT) [11] (http://www.eventhorizontelescope.org), which in the near future will join all largest ground-based submillimeter telescopes and observatories in a single VLBI network.

The present paper is prepared as a result of discussions of scientific objectives of Millimetron at scientific seminars and several symposia.\footnote{See the report by V.N. Lukash and I.D. Novikov at Groningen (2013) (\url{https://streaming1.service.rug.nl/p2player/player.aspx?path=
cit_mobiel/2013/04/12/3/video_post.wmv&mediatyoe=recordings}.}

The Sections of the paper correspond to different astrophysical topics, where Millimetron can significantly contribute. The interferometric regime of observations is assumed to be used in tasks formulated in Sections 2.4, 4.2, 4.5, 5.1-5.5, 7.2; to solve tasks described in other sections of this paper, single-mirror regime of observations can be used. It should be noted that the list of scientific tasks presented in this paper is preliminary, and from the broad scientific community we are waiting for both new unique scientific tasks and more detailed elaboration of the proposed ones. In June 2014, principal scientific tasks of the Millimetron observatory were discussed at an international symposium in Paris \footnote{http://workshop.asc.rssi.ru}. Using the results of this symposium and based on the preliminary list of scientific problems presented in this paper, the scientific program of the Millimetron observatory will be prepared.

\section{Interstellar medium and star formation regions in the Galaxy}
\subsection{Structure and kinematics of the interstellar medium}
Currently, the problem of star formation [12], as well as its relation to the general evolution of the interstellar medium (ISM), is an important astrophysical issue. Observational data suggest that the birthrate of new stellar and planetary systems and their parameters are determined by the basic properties of ISM: its structure, kinematics, pressure, temperature, magnetic field, and matter returned back to ISM from evolved stars. The galaxy environment effects (accretion and ram pressure of intergalactic matter, interaction with other galaxies) can also play an important role.

Global processes of star formation imply that it is necessary to investigate it in a general context of structure, kinematics and evolution of the interstellar medium using as broad sample of objects as possible. The densest ISM regions, where star formation occurs have a low temperature and, therefore, require observations in the FIR and submillimeter ranges, which can be done only from space. For a deeper study, a high sensitivity is also necessary, resulting in the requirement of large-size telescope mirrors, cooling, as well as increased demands for the detector parameters. The Millimetron project satisfies these requirements.

The most promising targets to be observed with Millimetron include cold ($10-20$ K) gas and dust clumps, which are difficult to detect with less sensitive instruments, ``hot cores'' and high-speed bipolar outflows, diffuse clouds, submillimeter masers, ISM in other galaxies.

Millimetron will study general characteristics of the interstellar medium in various galaxies, statistical properties of dense condensations, structure and kinematics of interstellar clouds, the earliest stages of star formation, mechanisms of massive star formation, structure and properties of circumstellar shells and planetary nebulae, synthesis and proliferation of different molecules in ISM including complex organic molecules.

High sensitivity of Millimetron in the single-dish mode will allow observing individual clouds with a characteristic temperature of about 20 K and a mass of order of one solar mass ($M_\odot$) at a distance up to 1 Mpc. The molecular cloud complex Sgr B2, where many molecules were found playing an important role in cooling and condensation of clouds, including the molecular ion H$_3$O$^+$, can provide an example [13]. This ion, which decays into water or hydroxyl due to dissociative recombination, is very important to an overall understanding of the chemistry of oxygen in the interstellar medium.

The submillimeter radiation from molecules and atoms arises at much lower temperatures than in the visible and infrared ranges. This means that by analyzing the submillimeter data, one can examine the cold ISM, in particular probe the content of {\it hidden hydrogen} [14]. The most important spectral lines of these observations include HD transitions at a wavelength of 112 microns and [CII] at a wavelength of 158 microns. A very interesting target for ISM research is a molecule HeH$^+$ (transition $J=1-0$ at a wavelength of 149 microns), which has not yet been detected in space. Conditions of formation and excitation for this molecule are substantially different from those for the majority of interstellar molecules. Models (e.g., [15, 16]) indicate that HeH$^+$ should be especially abundant near the sources of extreme ultraviolet (UV) and X-ray radiation.

An important area of research is a study of the ISM diffuse component by absorption lines of different molecules at submillimeter wavelengths. The Herschel space telescope has already demonstrated great possibilities of such researches. In particular, they allow determining the rate of ionization by cosmic rays in various regions (OH$^+$, H$_2$O$^+$ and H$_3$O$^+$ lines), turbulence dissipation rate (CH$^+$ and  SH$^+$ lines), and total distribution of molecular hydrogen (HD line). Such measurements require a sufficiently bright background source. For Millimetron with its much greater collecting area, the number of such sources will be larger than for the Herschel space telescope, which opens up the possibility of a much more complete coverage of the galactic plane.

Another source of information about the structure and kinematics of the interstellar medium are polarization measurements, which help to explore structure of magnetic fields in star-forming regions.

Ground-based observations of polarization in the FIR and submillimeter ranges can be carried out only in certain transparency windows. The possibility of constructing of the so-called polarization curve, i.e. the dependence of the polarization efficiency on wavelength, would allow us to explain not only the structure of magnetic field in the star-forming regions, but also a mechanism of dust particles orientation (Fig. 2).

\begin{figure}
\begin{center}
\includegraphics[width=7.5cm]{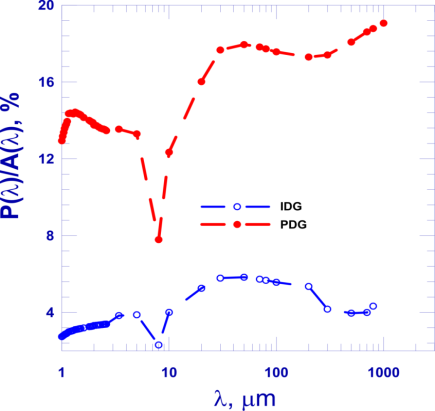}
\end{center}
\caption{The polarization efficiency curve for Perfect Davies Greenstein (PDG) (the upper curve) and Imperfect Davies Greenstein (IDG) (the bottom curve) orientation of interstellar dust grains [17]; $P$ is the polarization degree at the wavelength $\lambda$ in per cents, $A(\lambda)$ is the interstellar absorption at the wavelength $\lambda$.}
\end{figure}

Observations of dust emission in the submillimeter range are an important source of information about the star formation. The modern data obtained by the Herschel space telescope show that star formation occurs in thin (<0.1 pc) gas-dust filaments. Parameters of these filaments are not fully determined, and in particular a role of magnetic field in their formation is not clarified. To answer this question, we need higher angular resolution and sensitivity (better than that of the Herschel telescope), and availability of polarization measurements. The parameters of the space observatory Millimetron satisfy these requirements. In addition, contribution of Millimetron to star formation and evolution studies is enhanced by the possibility to observe more distant star forming regions, including extragalactic ones. In recent years, a new paradigm (where the primary role is given to large-scale star formation complexes [18-22]) is developing based on observational data obtained by the Herschel telescope and new theoretical researches. The heuristic role of this new paradigm is extremely important because it relates a large variety of phenomena: from the spiral arm and interarm flows several kiloparsecs in size to cores of molecular clouds and protostellar condensations as small as several astronomical units (AU). However, many aspects of the star formation process remain unclear and poorly investigated. Observations of IR lines at a wavelength of 158 microns show that the galactic disk contains a large amount of CO-dark molecular gas [23]. Theoretical studies of the formation of gas clouds in the Galaxy also show that temperature and density of molecular clouds vary strongly (see, e.g., the phase diagram in paper [22]), which may indicate existence of large masses of gas mostly consisting of molecular hydrogen with rather small abundance of CO molecules. Distribution and parameters of this gas have been studied only in a narrow band of the Galactic plane, and only in selected directions. Therefore, the measurements of the relative velocities and positions of gas clouds emitting in [CII] lines at a wavelength of 158 microns, as well as of submillimeter and millimeter radiation of dust, are needed to determine mechanisms of formation of star formation complexes, since such measurements form the basis of studies of morphology, kinematics and evolution of these large-scale objects.

Collective outflows of matter from the forming star clusters are another virtually unexplored important type of motion in star-forming regions. It is known that all star clusters eventually get rid of their parent gas. Clumps of molecular gas in the direction of forming star clusters in star-forming region S235 are likely to represent such collective outflows [24, 25].  A deeper understanding of the collective outflows from forming clusters requires observations of CO-dark molecular gas in the [CII] lines at wavelength of 158 microns, which can be effectively carried out by Millimetron.

Studies of objects with large angular dimensions should be preferably carried out step by step. During the initial stage it is necessary to determine an overall distribution of the radiation in the test line with a matrix spectrometer in the broadband spectroscopy mode with an average spectral resolution. To study a detailed kinematics, a high-resolution spectrometer is proposed to be used in a number of key areas identified in the first phase of the Millimetron observations or using data obtained previously by other instruments.

\subsection{Star formation regions in the Galaxy}
Spectroscopic observations with high resolution will allow a detailed study of the molecular structure of protostellar objects. Here of particular importance are studies in the short-wavelength range of Millimetron at wavelengths less than 300 microns. The number of lines in this spectral range is smaller than in the longer wavelength region [26], which facilitates both the identification of the lines and their analysis (Fig. 3). The number of lines is still very large, including both simple compounds lines, and lines from complex organic molecules which are interesting from the astrobiology viewpoint. Some line transitions, available for Millimetron observations, are presented in Table 3.

\begin{figure}
\begin{center}
\includegraphics[width=7.5cm]{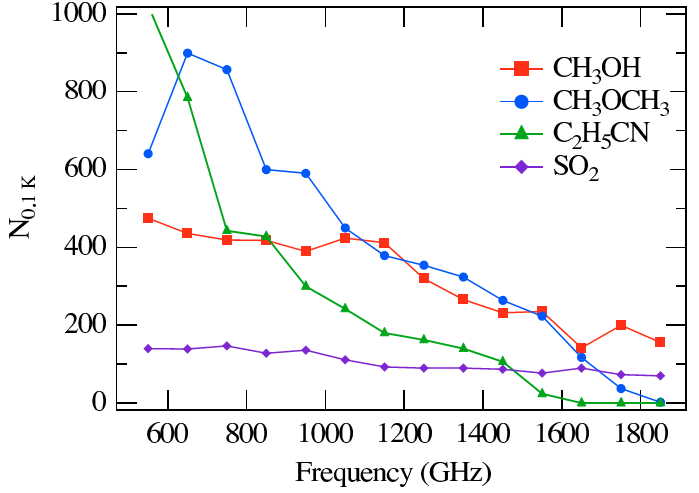}
\end{center}
\caption{The predicted number $N_{0,1K}$ of some molecular lines  with maximum emission at a temperature of above 0.1 K as a function of frequency [26].}
\end{figure}

The high spectral resolution  ($\sim10^6$) and high sensitivity achievable in the  Millimetron project in observations of molecular lines make it possible to characterize the physical parameters and motion of gas in the protostellar objects.

\begin{table}
\caption{Important submillimeter atomic and molecular transitions}
\begin{center}
\begin{tabular}{|m{3.166cm}|m{8.637cm}|}
\hline
Species &
Frequencies (GHz)\\\hline
C I &
492, 809\\\hline
O I &
2060, 4745\\\hline
HD &
2675, 5332\\\hline
OH &
1835, 2510, 3789, ...\\\hline
CH &
537, 1477, 1657, 2007, 2011, 4056, 4071, ...\\\hline
HF &
1232, 2463\\\hline
H$_2$O &
557, 988, 1113, 1670, 2774, 2969, ...\\\hline
HDO &
465, 894\\\hline
C II &
1901\\\hline
N II &
1461, 2459\\\hline
O III &
3393, 5787\\\hline
N III &
5230\\\hline
HeH$^+$ &
2010, 4009\\\hline
OH$^+$ &
972, 1033, 1960, ...\\\hline
CH$^+$ &
835, 1669, ...\\\hline
SH$^+$ &
526, 683, 893, 1050, ...\\\hline
H$_2$O$^+$  &
1115, 1140, ...\\\hline
H$_3$O$^+$  &
985, 1656, ...\\\hline
H$_3^+$ &
3150\\\hline
H$_2$D$^+$ &
1370, 2577\\\hline
D$_2$H$^+$ &
1477\\\hline
\end{tabular}
\end{center}
\end{table}

One of the most interesting lines shown in Table 3 belongs to atomic oxygen. The OI line at a wavelength of 63 microns significantly contributes to cooling of the warm ISM and photodissociation regions with high density and intense UV radiation. Observations of this line are needed to study the energy balance in ISM, but none of the existing instruments can observe this line. Observations of the 63 $\mu$m line by Millimetron will provide information about the content of oxygen in the interstellar medium, as well as will help to solve the problem of the intensity ratio of the oxygen lines at wavelengths 63 and 145 microns detected in observations by ISO (Infrared Space Observatory) [27]. Joint observations of oxygen, water, molecular oxygen and hydroxyl lines will help to explain the chemical evolution of oxygen compounds.

The rate of accretion onto protostellar objects is one of the key questions in protostellar evolution studies. As direct measurements of the accretion rate are very difficult, it is usually inferred from molecular outflow parameters [28], for example, using CO observations.  However, to understand the transition from the outflow rate to the accretion rate more information is required: the velocity and extent of the outflow, the inclination of the system, parameters of the surrounding material. Observations of the 63 $\mu$m[OI] line will measure the accretion rate in a more direct way, but this will also require a high angular and high spectral resolution.

Another way of probing the accreting matter is to observe absorption lines, which would guarantee that the absorbing material is in front of a growing protostar. Recently it was shown [29] that a promising line in this respect could be an ammonia line having the wavelength of 166 microns. FIR spectral observations by Millimetron will provide an opportunity to study accretion onto protostars with a spatial resolution four times higher than the SOFIA telescope. This gain in spatial resolution can be critical for star formation studies.

Observations of neutral oxygen and ionized carbon lines with high angular and high spatial resolution are valuable to probe the evolution of ionized hydrogen regions. In particular, a numerical simulation shows that the width of the ionized carbon region around a young massive star significantly depends on the parameters of the star and the surrounding gas density [30].

One of the tasks of the Millimetron operations will be high-resolution spectral surveys covering frequency bands of a few tens or hundreds GHz. The Millimetron high sensitivity and lack of atmospheric absorption will enable surveying faint and hence poorly studied sources (for example, ``hot corinos'', that is, hot regions near low-mass protostars). As a result, in addition to determining the main physical parameters and molecular composition of these sources, new molecules can be detected, including those important from the viewpoint of astrobiology.

One of the tasks for Millimetron space observatory will be studying the cosmic masers in the millimeter and submillimeter wavelengths in the single-dish mode. Bright masers occur in water vapour line $6_{1,6}-5_{2,3}$ at a frequency of 22 GHz. However, there are other known maser lines of H$_2$O in the submillimeter wavelength range. Submillimeter lines of water vapour are difficult or even impossible to observe from Earth due to strong absorption in the atmosphere, so observations of H$_2$O masers are carried out almost exclusively at frequency 22 GHz. As a result, even very crude model of these objects cannot be built. Some progress in  observations of submillimeter maser lines have been made at high-altitude astronomical observatories and using space-based observatories SWAS (Submillimeter Wave Astronomy Satellite), Odin, and Herschel. Further progress will be possible after the launch of Millimetron, which will be the best submillimeter maser observatory among space observatories to be launched within the next  10-15 years.

\subsection{Protoplanetary disks and protostellar objects}
One of the key issues in the physics of ProtoPlanetary disks (PPD) is their mass. Presently, to determine the PPD masses, mainly millimeter dust emission  observations are used, under the assumption that the dust is well mixed with  gas. However, the sensitivity of modern telescopes is insufficient for detection of low-mass and distant PPD. There are many disks, where only upper limits of radiation fluxes (<10 mJy) are obtained. Meanwhile, only the accurate disk mass evaluation can clarify whether the disk is able to form a planetary system. It would be highly desirable to find a direct indicator of the gas mass. Presumably, HD molecule radiation (Fig. 4) at 112 and 56 microns [31] could be such an indicator.

One of the biggest challenges for Millimetron may be observations of water in PPD. In these objects, both a ``warm'' (close to the star) and ``cold'' (more distant) water reservoirs are possible. At present, water is commonly viewed as being the main factor determining the structure of planetary systems (the ``snow line''). Observations of water lines on the Herschel space telescope were carried out only for a few PPD, and observations of the ``cold'' water in them yielded conflicting results [32, 33]. Obviously, to clarify the role of water in the formation of planetary systems a more significant sample is needed, which requires instruments with greater sensitivity and better angular resolution. The combination of water lines in various parts of the Millimetron range allows to make conclusions about the spatial distribution of water in the disk, in particular, about actual location of the ``snow line''.

In PPDs, observations of molecular oxygen and ro-vibrational lines of complex organic and simple compounds in the planetary formation zones as well as of less common isomers of previously detected molecules (or discovered by ALMA) will also be feasible. High temperatures (over 100 K) in planetary formation zones at distances less than 5--20 AU from the star result in populating high transitions, especially in complex molecules. An example is provided by the detection of organic molecules in the inner parts of the disks with the Spitzer telescope (see, for example, [34]). Calculations of the PPD chemical structure show [35] that in the planetary formation zones, column densities of such molecules as methyl cyanide, formic acid, etc., with lines in the Millimetron range, reach large values.

In the submillimeter and far-infrared (100 $\mu$m) range, observations of the large particles forming in PPD are possible. It is very important to calibrate receivers properly, with a wide and continuous spectral coverage. Figure 5a shows theoretical spectra of a PPD in the globule CB26 in comparison with the Herschel telescope observations. Error bars indicate the flux calibration uncertainty. The figure shows that in order to find the mass distribution and size of dust particles, high-precision observations of disks at wavelengths of about 100 microns are needed. To solve this problem the single-mirror mode is sufficient.

FIR spectroscopy with a moderate spectral resolution allows the bright CO and water to be detected, as well as the parameters of the PPD inner regions to be determined.

\begin{figure}
\begin{center}
\includegraphics[width=7.5cm]{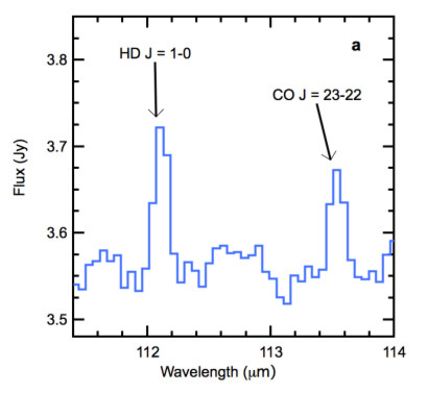}
\end{center}
\caption{The HD molecule line in the spectrum of protoplanetary disk around TW Hya (http://cosmos.esa.int/web/Herschel/home).}
\end{figure}

In addition to the PPD studies, searching and studying of cold gas and dust clouds in the Galaxy is an important task. One of the main sources of information about the physical conditions in the prestellar and protostellar objects is their broadband spectral energy distribution (SED). It was SED that became the basis for the currently accepted classification system of these objects. Moreover, for objects of class -1 (prestellar core) and 0 (the earliest stage of evolution in the presence of a central IR source), the spectral maximum falls in the FIR and submillimeter range. A space telescope will allow building SED of prestellar cores without breaks caused by atmospheric transparency windows. The detailed shape of the spectrum will clarify the evolutionary status of a specific core. Until now, the identification of a core as being the pre- or protostellar one has been based on the absence or presence of a compact source in the core. The relative number of pre-stellar and protostellar objects forms the basis for estimation of the relative duration of the corresponding evolutionary stages. However, for example, observations of the low-mass core L1014 by the Spitzer space infrared telescope revealed the presence of a weak compact internal source by excess radiation at wavelengths of less than 70 microns [37].

Pavlyuchenkov et al. [38] show the existence of such a problem also for massive cores. Paper [37] analyzed observations of two massive cores of infrared dark clouds (IRDC). Near-IR and millimeter studies classified these cores as starless. However, analysis of the spectrum at a wavelength of 70 micrometers revealed that in both cases the cores already hide embedded compact sources of radiation --- protostars.

The reasons for the importance of the 50-150 micron range are illustrated in Fig. 5b, which shows the result of fitting of the near-IR spectrum of a typical protostellar object with a model from [39, 40].  The formal best-fit model is shown by the solid black curve. However, similar fits can be obtained by other models with the object mass ranging from $1M_\odot$ to $10M_\odot$. Figure 5 clearly shows that the maximum emission of a typical protostellar object falls in the FIR range. Inclusion in the model of the same object the data on the radiation at a wavelength of 70 microns reduces the mass uncertainty by three times.

\begin{figure}
\begin{center}
\includegraphics[width=7.5cm]{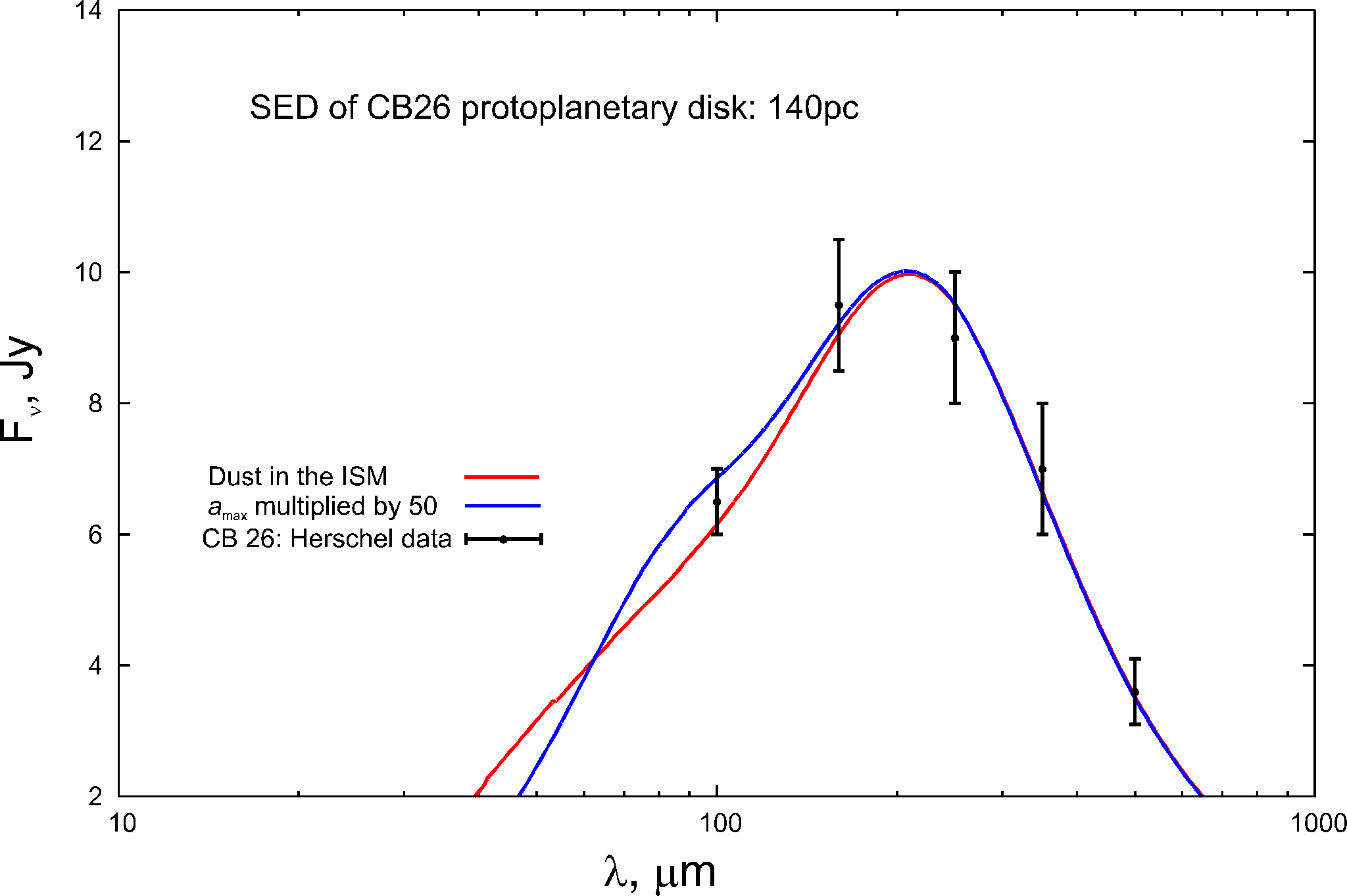}
\includegraphics[width=7.5cm]{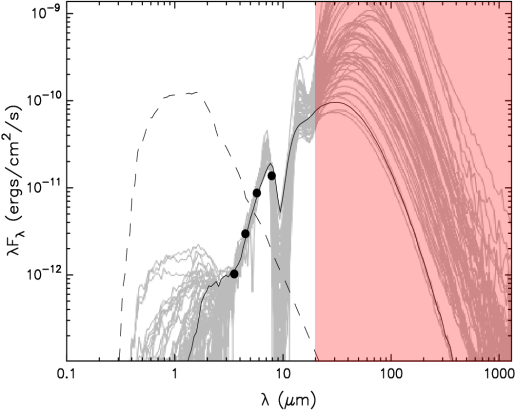}
\end{center}
\caption{(a) Spectral energy distribution (SED)  of the protoplanetary disk in globule CB26 observed by the Herschel telescope. The red curve is a model of the dust with typical ISM parameters (the maximum grain size $a_{max}=0.25$ $\mu$m ), the blue curve is a model for $a_{max}$ 50 times as large as the typical value (calculations according to the model presented in paper [36]).  (b) The fitting of the observed near IR spectrum (filled circles) from the typical protostellar object. The formal best-fit solution is shown by the thick black curve. The grey curves show spectra of objects with masses from $1M_\odot$ to 10 $M_\odot$, which equally well fit the observed near IR data but significantly deviate in FIR (http://caravan.astro.wisc.edu/protostars/). The dashed curve shows the best-fit solution for the photosphere emission from the central source.}
\end{figure}

The 50-150 micron range was available for observations by the Herschel space telescope. However, the lack of high angular resolution of the instrument did not allow a detailed investigation of the protostellar objects at large distances, which significantly limits their sample. The higher angular resolution of Millimetron enables observations at longer wavelengths. In particular, using the  PACS (Photodetector Array Camera and Spectrometer) detector of  the Herschel telescope (70, 100 and 160 microns) many point-like objects  inside many IRDC were discovered  [42]. However, at longer wavelengths (250-350 microns) such point-like objects were not resolved. Meanwhile, it is necessary to determine temperature, mass and luminosity of these sources more accurately. The high sensitivity of Millimetron will help to discover more compact protostellar sources by including into the statistics distant and (or) denser objects.

In addition to cold gas and dust clouds in star-forming regions,  hotter objects have been detected. They show a dust continuum and numerous spectral lines of gas-phase molecules. The spectral energy distribution in such objects is still poorly explored, but is a very important task.

Another important field of research is the study of high-speed bipolar outflows formed by accretion disks around protostars and young stars (for example, the object S255IR that emits 1.1 mm continuum) (Fig. 6). By the appearance, a bipolar outflow is similar to a black hole jet, which suggests a possible similarity in physical mechanisms of both phenomena. The origin of bipolar outflows is uncertain yet, so their study is an actual task.

\begin{figure}
\begin{center}
\includegraphics[width=7.5cm]{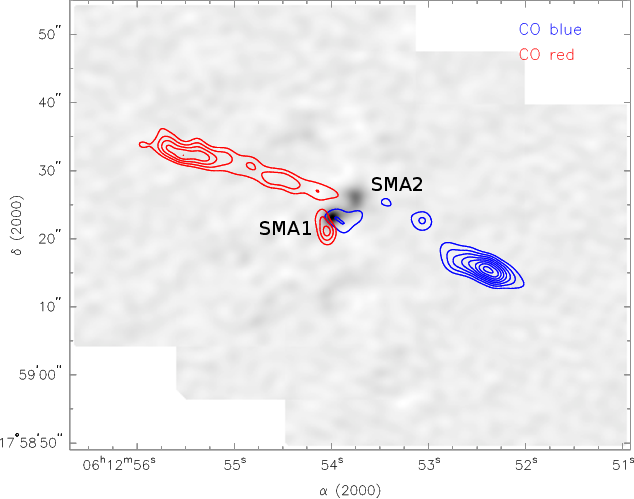}
\end{center}
\caption{Example of q high-velocity bipolar outflow. Blue and red contours show the emission in the blue and red wing of the CO(3-2) line, respectively. Sky coordinates $\alpha,\; \delta$ (epoch J2000) [41].}
\end{figure}

As a related problem, we can also mention submillimeter observations of asteroids and comets. Observations of asteroids in the reflected (scattered) light could not give reliable information about their size, because their albedo should be assumed. Observations of the intrinsic radiation of asteroids in this respect are more reliable. In addition, the emission spectra of asteroids allow us not only to evaluate their size, but also to obtain information on the chemical composition and surface structure [43]. However, the temperature of asteroids (especially away from the Sun) is low, and so their emission fall in the infrared and submillimeter range, which requires space observations. Spectral observations of comets make it possible to clarify their molecular composition and obtain information about the evolution of matter in the early Solar System.

\subsection{Maser sources}
Masers quantum transitions in molecules are a powerful tool to study different astrophysical sources, such as accretion disks around supermassive black holes in galactic cores, proto-stellar / proto-planetary disks and outflows from young stars in star-forming regions, regions of interaction between expanding HII regions and supernova remnants and dense surrounding gas clumps, expanding shells and jets associated with evolved star [44, 45]. Observations of masers are widely used to detect sources with unique physical and evolutionary status, to measure their distances, to study their parameters, kinematics and structure of accretion disks [46-48]. Modern scientific projects RadioAstron, BeSSeL (Bar and Spiral Structure Legacy Survey), MCP (Megamaser Cosmology Progect), and MMB (Methanol Multibeam Survey) demonstrate high possibilities of maser observations as a research tool. The value of this instrument is determined by a precise measurement of positions of the objects. Maser sources have a small angular size, so observations in interferometry mode are most important.

In the ground-Space VLBI (SVLBI) mode the receiving equipment of Millimetron allows observations of water masers at the transition frequency 22235.08 MHz. The possibility of maser observations in the SVLBI mode was shown by the RadioAstron space mission.

The scientific program to study cosmic masers using ground-space interferometer RadioAstron included observations of 19 water molecule maser sources. These masers are very compact (often not even resolved using the largest ground-based bases) objects that have the highest brightness temperature, sometimes exceeding $10^{17}$ K [49]. Because of these properties, masers can be used to high-precision studies of the kinematics and physical parameters of objects in our and other galaxies. Observations by the ground-space interferometer allow to resolve the most compact components and to evaluate the brightness temperature of the maser source and its size, which is necessary to clarify the pumping mechanism and to model the emitting region. During observations of the water line at the frequency of 22 GHz, radiation from extremely compact maser components in the directions to the four star formation was detected: Orion A, W31RS5, W51M/S and Cepheus A [50].

Some sources (Cepheus A) show the presence of very compact substructure maser details with a size of the order 10 micro arcseconds (which corresponds to 0.007 a.u.). Moreover, the objects are moving with a high relative velocity. This means that the maser source has a complex spatial and kinematic hyperfine structure on scales comparable to the size of the Sun. This picture is probably an indication that in this case the maser emission arises from the proto-stellar / proto-planetary disks or the smallest turbulence cells corresponding to the dissipation scale. For disks of all types the most important and still outstanding issue is the mechanism of the angular momentum transfer [51]. The turbulent viscosity is conventionally considered to be such a mechanism, but as yet there is no consensus on the mechanism of the turbulence excitation [51, 52].

The Millimetron equipment allows observing masers in the submillimeter range, as well as masers formed in other galaxies and in evolved stars, which requires better sensitivity than that of the RadioAstron. At the moment, the possibility of observing in the SVLBI mode is confirmed only for of masers in star-forming regions of the Galaxy at hydroxyl and water molecule centimeter transitions.

\section{Stars and planets}
\subsection{Direct observations of exoplanets}
One of the promising methods for studying extrasolar planets is their direct observations.

In the single-dish mode Millimetron is able to observe massive planets far remote from central stars. The more distant the system, the larger should be the orbital radius of the planet to be resolved by a telescope. Unfortunately, Earth-like planets at such distances have very low radiation intensities, so only gas giants can be observable in this mode. Table 4 lists extrasolar planets (as of the end of 2013), which will be available for direct Millimetron observation. Only three planets from this list could be observed by the Herschel telescope, and in addition planet Fomalhaut b was observed at the limit of sensitivity. Two planets shown in Table 4 were discovered already after the Herschel telescope operation had been completed. Note that in the range of Millimetron measurements, the star and a remote gas giant have the lowest contrast, which facilitates their observation [53].

Observations of known transiting planets are of particular interest. Observations of planetary transits and antitransits allow the orbital parameters of the planet, its mass and radius to be determined, as well as the absorption spectra of the upper layers of its atmosphere to be obtained. Since 2003, there has been a series of detailed spectroscopic observations of transits and antitransits of different types of exoplanets [54-58]. The high accuracy of Millimetron enables us to carry out spectroscopic observations of transits and antitransits and thus to obtain unprecedentedly comprehensive information about these systems.

\begin{table}
\begin{center}
\caption{Exoplanets which can be directly observed with Millimetron}
\begin{tabular}{|l|l|m{1.5cm}|m{1.5cm}|l|l|m{2cm}|l|}
\hline
 Name &
 Mass, $M_\odot$ &
 semiaxis, au &
 distance, pc &
 stellar type &
 age &
 effective temperature &
 <<Herschel>>\\\hline
 Fomalhaut b &
 3 &
 115 &
 7.704 &
 A3 V &
 0.44 &
 8590 &
 limit\\\hline
 HN Peg b &
 16 &
 795 &
 18.4 &
 G0 V &
 0.2 &
 {}- &
 yes\\\hline
 WD 0806-661B &
 8 &
 2500 &
 19.2 &
 DQ D &
 1.5 &
 {}- &
 yes\\\hline
 AB Pic b &
 13.5 &
 275 &
 47.3 &
 K2 V &
 0.03 &
 4875 &
 no\\\hline
 Ross 458 (AB) &
 8.5 &
 1168 &
 114 &
 M2 V &
 0.475 &
 {}- &
 no\\\hline
 SR 12 AB c &
 13 &
 1083 &
 125 &
 K4-M2 &
 0.001 &
 {}- &
 no\\\hline
 FU Tau b &
 15 &
 800 &
 140 &
 M7.25 &
 0.001 &
 2838 &
 no\\\hline
 U Sco CTIO 108 &
 16 &
 670 &
 145 &
 M7 &
 0.011 &
 2600 &
 no\\\hline
 HIP 78530 b &
 23 &
 710 &
 156.7 &
 B9 V &
 0.011 &
 10500 &
 no\\\hline
 GU Psc b &
 11 &
 2000 &
 48 &
 M3 &
 0.1 &
 {}- &
 {}-\\\hline
 HD 106906b &
 11 &
 654 &
 92 &
 F5V &
 0.013 &
 6516 &
 {}-\\\hline
\end{tabular}
\end{center}
\end{table}

\subsection{Mass loss at late stages of stellar evolution}

At the late evolutionary stages, low- and intermediate mass stars (which include the Sun) experience an intensive mass loss. This is manifested in an expansion of the stellar envelope with the subsequent transformation into a planetary nebula. Studies of circumstellar medium during and after the asymptotic giant branch (AGB) are also closely related to the ISM enrichment by heavy elements.

High abilities of space investigation of low- and intermediate mass stars in the infrared range were demonstrated by such key programs of the Herschel telescope as HIFISTARS, ``The circumstellar environment in post-main-sequence objects'', and others \footnote{\url{http://Herschel.cf.ac.uk/key-programmes/stars}.}. Higher angular resolution, sensitivity, and a wide range of Millimetron not only can significantly improve the quality of research evolved stars, but also can formulate a number of new challenging tasks in the study of these objects.

\begin{figure}
\begin{center}
\includegraphics[width=15cm]{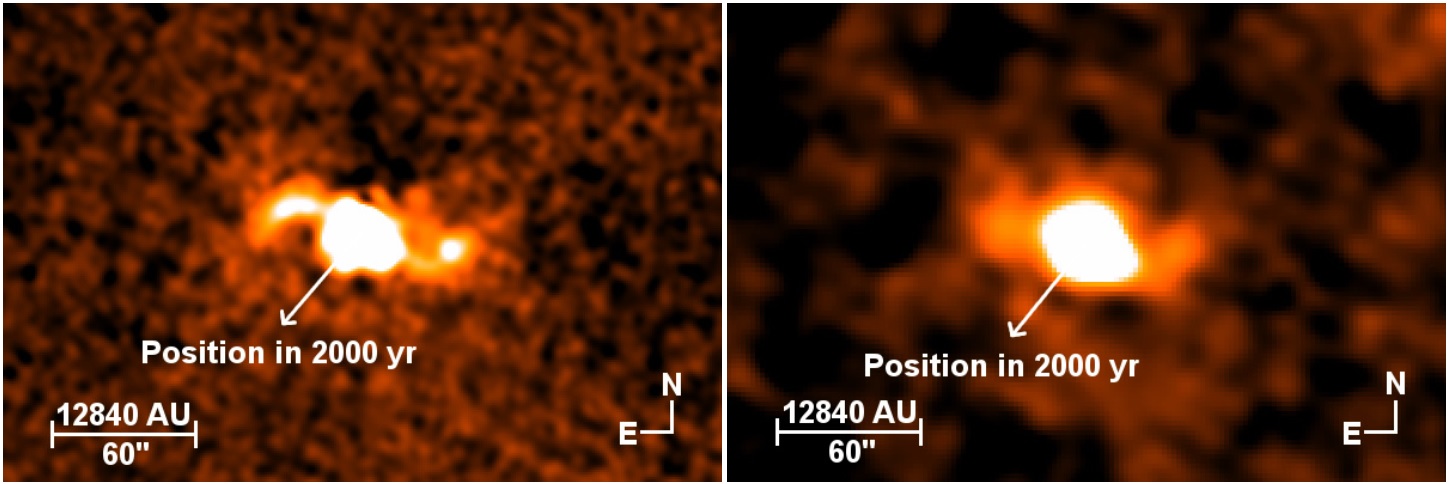}
\end{center}
\caption{Image of the star R Aqr in the continuum at wavelength 70 $\mu$m  (a) and 160 $\mu$m  (b). Arrows show the proper motion of the object [59].}
\end{figure}

Observations from the high-resolution spectrometer will allow, in particular, the following studies.
\begin{itemize}
\item Kinematics of outflows from AGB stars. Millimetron allows observations of the most highly excited molecular lines of CO, H$_2$O, and other lines formed close to the stars. Such observations are impossible with other instruments, including the Herschel telescope, and will help to answer a number of important issues related to the outflow mechanism. For example, according to modern concepts the radiation pressure on the dust in oxygen stars is not large enough to explain the observed mass loss rates.
\item The cause of asymmetry of the stellar shells, symmetrical on the AGB stage and asymmetric at the stage of proto-planetary and planetary nebulae. Currently, there are several hypotheses. One of them is related to the binarity of the star [59]. To confirm or refute this hypothesis observations with high angular resolution are required, which are possible by  Millimetron.
\item The composition and physical parameters of the interacting regions of shells of moving evolved stars with an environment. Observations made by the Herschel telescope revealed the presence of complex structures and motions in such objects which allows us to investigate the stellar envelope, circumstellar medium, and motion of the stars [59, 60] (see Figure 7).
\item The formation of molecules and dust in the evolved stars shells [61, 62].
The use of the short-wavelength matrix spectrometer will allow, in particular, to carry out the following studies:
\item Observations of extended planetary nebulae in the [NIII], [OIII], [OII], [CII] lines,  etc., to study the dependence of the chemical composition and physical parameters of the gas on the distance to the star. The methodology that was developed in paper [63], will be further improved to analyze the Millimetron data with a better sensitivity. These studies are related, in particular, to the problem of the origin of poor hydrogen stars;
\item Observations of clumps and other irregularities in the structure of extended planetary nebulae to study variations of temperature and density, which largely determine the chemical composition and evolution of these objects;
\item Studies of the structure of stellar shells interacting with the environment. These studies also require observations using matrix instruments.
\end{itemize}

\subsection{Searches for extraterrestrial life}
Due to the absence of atmosphere, the Millimetron observatory will be capable of studying many spectral lines of water molecules and more complex organic compounds which are difficult or impossible to observe from the Earth surface. The study of these lines in the solar system and PPD will help to determine the origin and primary molecular composition of the Earth oceans, as well as to draw conclusions about the abundance of planets containing liquid water on the surface, and therefore having suitable conditions for life.

For years, the search for manifestations of extraterrestrial civilizations is one of the most ambitious projects of the mankind. Major efforts are now focused on interception of messages from extraterrestrial civilizations and the millimeter range is promising for these purposes. Paper [64] justifies the benefits of this range for the directed transmission in the midsection of the cosmic microwave background. A characteristic marker of this region of the spectrum can be the line of positronium hyperfine splitting at 203 GHz ($\lambda=1.5$ mm), an analogue of the 21 cm line of hydrogen atom. Preliminary observations have already begun [65]. Search for positronium using Millimetron is an independent important task.

Along with the search for signals from extraterrestrial intelligence, traces of astro-engineering activities are being searched for. In particular, a well-developed civilization is able to surround a star by a system of structures, intercepting and using a significant portion of stellar energy (the so-called Dyson sphere [66]) which should re-emit the whole or part of the energy at lower frequencies than a radiation frequency of the star itself. For the Sun and the Dyson sphere with radius of 1 AU the temperature of the sphere will be about 300 K. It can be expected that the use of more advanced technologies will be associated with the use of low temperatures and the position of the emission maximum will shift from 20 microns towards longer wavelengths. Therefore, such objects should be searched for most effectively in the infrared range up to the wavelength corresponding to the maximum of space radiation (1.5 mm). The first sky FIR survey aimed at the detection and  spectral measurements of astronomical objects was carried out on the IRAS satellite (InfraRed Astronomical Satellite). 250 thousand point-like sources were found. Results of the search for objects similar to the Dyson sphere, are reported in papers [67, 68] (Fig. 8); several objects were found, where a natural origin has not been still reliably proven.

\begin{figure}
\begin{center}
\includegraphics[width=10.5cm]{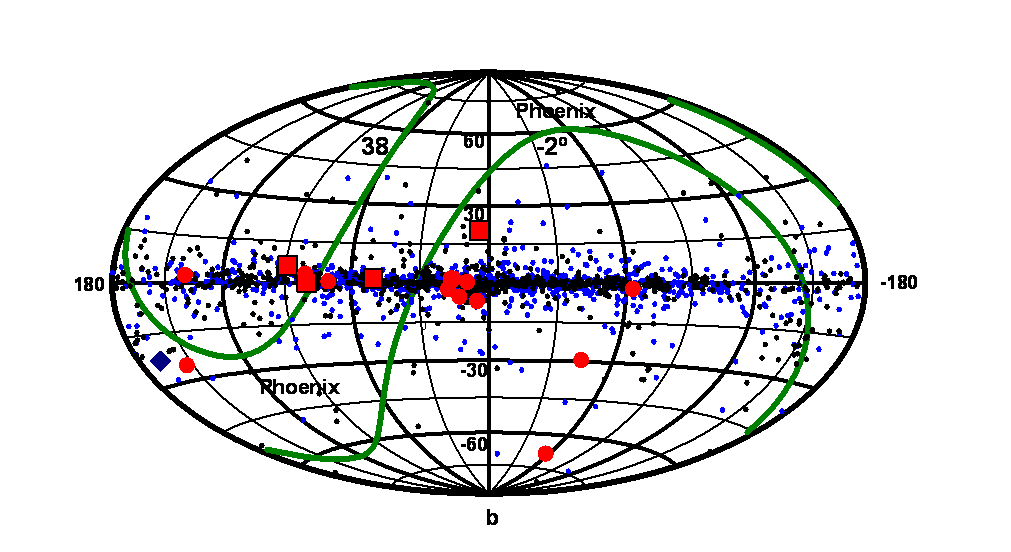}
\end{center}
\caption{Sky location of 16 possible Dyson sphere candidates (red symbols). Three of them (red squares) show the least deviation from a black body spectrum. The blue symbols show 2240 candidates selected from the IRAS catalogue. The green curves limit the sky region accessible for the Arecibo radio telescope participating in the SETI program [68].}
\end{figure}

Important criteria are the spectral parameters and their comparison with the black body spectra. The spectral maximum determines the temperature. The spectral index in the long-wavelength part of a power-law spectrum is -2 for the black body and -3 and -4 for amorphous and metal dust particles, respectively, whose size is much smaller than a wavelength. Temperature, flux, form of the red part of the spectrum and a distance from the source can be used to estimate the size of the source, and to distinguish it from natural clouds of dust or stones emitting in the infrared (protostellar objects, old stars). To develop a reliable criterion of the search for the Dyson spheres, it is necessary to investigate in detail properties of natural sources, which can be accomplished by Millimetron.

\section{Supernovae and supernova remnants}
\subsection{White dwarfs}
FIR photometric observations of nearby cold white dwarfs allow
determination of their atmospheric composition, which can be used to
obtain precise ages of the objects and of the Galaxy itself.

White dwarfs are, apparently, the most numerous sky objects. They
can be separated into two large groups: hot and cold white dwarfs. A
cold white dwarf represents an observable end stage of  the white
dwarf evolution, therefore an age estimate of cold white dwarf can
be used to determine the age of the Galactic disk and halo, as well
as of the nearby globular clusters. White dwarf cooling, which plays
the crucial role in the white dwarf evolution, is not yet fully
understood. It depends on the white dwarf atmosphere composition.
Therefore, to determine the temperature, luminosity and age of cold
white dwarfs their atmospheric composition should be known.

Analysis of mid-IR observations of nearby cold white dwarfs with an
effective temperature of less than 6000 K revealed that the maximum
of emission in this range is somewhat lower than predicted by models
that well reproduced the observed luminosity function at wavelengths
shorter than 1 $\mu$m [69]. This situation is illustrated in Fig. 9,
where the observed spectral energy distribution in the wavelength
range from 0.1 to 100 $\mu$m is shown for cold white dwarf LHS 1126
together with model atmospheres of a different chemical composition.
Clearly, to solve the issue of atmospheric composition of cold white
dwarfs, high-precision FIR measurements are required.

The expected Rayleigh-Jeans fluxes from white dwarfs at distances closer than 100 pc are about a few mJy.

An interesting problem is to search for cold dust debris disks
around white dwarfs, which can be formed due to comet and asteroid
collisions, by performing single-dish FIR observations. For example,
white dwarf G2938 shows a FIR excess. In addition, in the atmosphere
of this object, as well as in other hot white dwarfs  heavy elements
were discovered, such as calcium and iron, which should have plunged
deeper into the star due to strong surface gravity [70]. A presence
of a debris disk irradiated by the white dwarf could be an
explanation. More than 20 white dwarfs surrounded by dust disks are
currently known from FIR observations. Presently, studies of white
dwarf atmospheres ``polluted'' by heavy elements are carried out in
the middle IR range: for example, photometric data was obtained by
WISE (Wide-Field Infrared Survey Explorer) [71]. Several white
dwarfs demonstrate an excess in this range [72].

\begin{figure}
\begin{center}
\includegraphics[width=7.5cm]{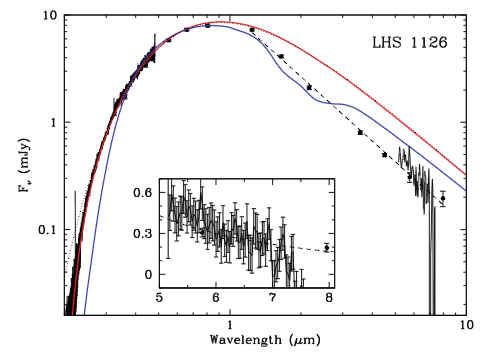}
\end{center}
\caption{Spectral energy distribution of the cold white dwarf LHS 1125 (model) [69]. The red and blue curves show models with different hydrogen to helium ratio in the atmosphere. Circles with error bard show the results of observations. The dashed curve, which best fits the observations, corresponds to the spectral slope --1.99. The inset zooms the spectrum at 5---8 $\mu$m.}
\end{figure}

In the last time, a presence of exoplanets around white dwarfs is
discussed. In principle, white dwarfs are sufficiently bright
objects to sustain water in the liquid state on surfaces of such
planets. The first planet was discovered around the star GI86, which
belongs to a binary system with a white dwarf. Both photometric and
spectroscopic studies of white dwarf atmospheres with heavy elements
and of white dwarfs surrounded by dust disks aimed at exploring
properties of the dust, atmospheric composition and discovering
exoplanetary systems around them, are important and interesting
tasks.

\subsection{Pulsar radio emission}

Spectra of most radio pulsars rapidly decrease with frequency [73]. However, some objects are observed in the Gigahertz range [74-76]. Here, there can be several new and interesting tasks.

In four radio pulsars (B 0329+54,  0355+54, 1929+10 and 2021+51) and in two anomalous X-ray pulsars (AXP) (XTE J 1810-197 and 1E1547-5408) spectral flattening is seen at several dozens of GHz (Fig. 10 a,b) or even an intensity increases when the frequency increases up to 87 GHz [75]. In radio pulsars and AXPx, the 43 GHz flux density lies in the range from 0.15 to 0.50 mJy [77]  and from 1 to 5 mJy [78, 79], respectively. The sensitivity of the bolometer for broad-band ($\gtrsim$ 10 GHz) observations in the 50-200 GHz frequency range with an exposure of several ten minutes is estimated to be around 0.1-0.5 mJy, which likely makes it possible to detect such pulsars.

\begin{figure}
\begin{center}
\includegraphics[width=7.5cm]{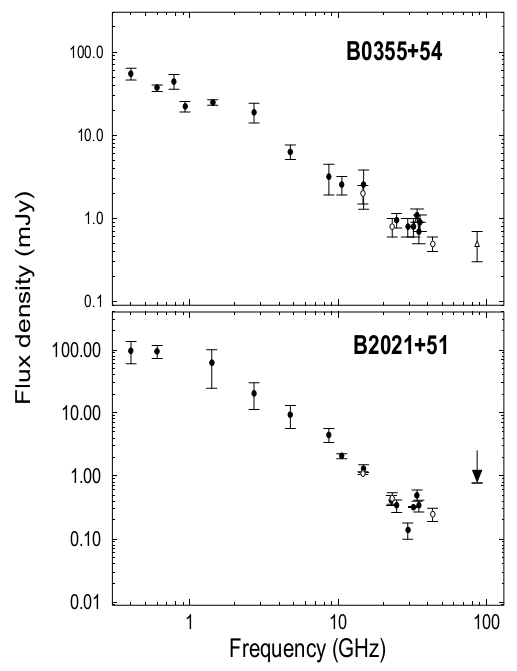}\\
\includegraphics[width=11.5cm]{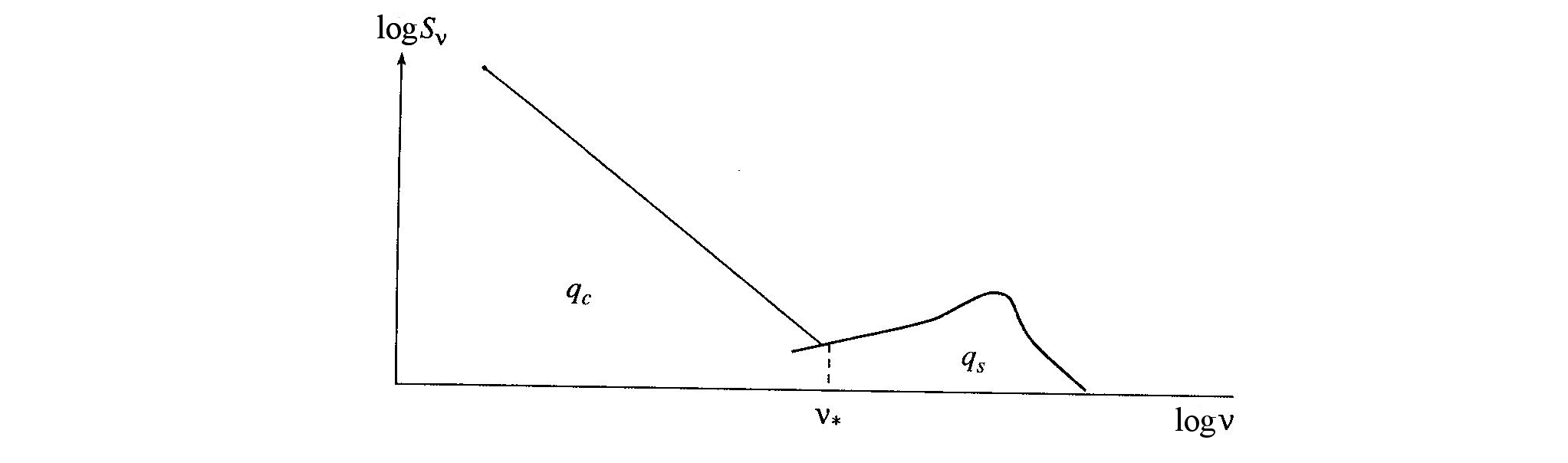}
\end{center}
\caption{Spectra of pulsars B0355+54 (a) and B2021+51 (b). Measurements at 87 MHz for B0355+54 and B2021+51 are shown by the open triangle and the arrow (the upper limit), respectively. (c) Theoretical spectrum of a radio pulsar: $S_\nu$ is the flux density, $q_c$ is the power emission generated due to cyclotron instability, $q_s$ is the power of synchrotron radiation.}
\end{figure}

Moreover, estimates show that in these objects magnetic field and
spin rotation axes misalignment can be small (less than 30$^\circ$). In
this case, the magnetosphere can extend far beyond the light
cylinder forming appreciable pitch-angles of relativistic electrons
(more than 0.01), and synchrotron radiation is generated [80, 81.
For typical parameters of these pulsars the intensity of the
radiation must increase with frequency starting from a few tens of
GHz. When the usual pulsar parameters are assumed the maximum
frequency is about $\nu_{max}\sim3\times 10^{11}$ Hz. A qualitative estimate of the millimeter
intensity increase is model-dependent. For a monoenergetic
 energy distribution the intensity increases as $\nu^{1/3}$,
suggesting twofold intensity increase at 275 GHz compared to 34 GHz.
For a power-law electron energy distribution and assuming a large
optical depth of the emitting region, the intensity increases as $\nu^{2.5}$,
and at 275 GHz the intensity increase can be as high as 200 times
(Fig. 10c). Testing these predictions is important to get deeper
insight into the models of magnetospheres of both radio pulsars and
AXPs.

Polarization measurements of pulsars with increased flux density at several tens of GHz play an important role. The degree of polarization of ordinary pulsars, as a rule, decreases with frequency. The switch-on of the synchrotron mechanism should increase the degree of polarization. Confirmation of this property would provide an additional argument supporting the synchrotron radiation hypothesis for pulsars. In addition, the pulse profile generated by the synchrotron mechanism differs from profiles expected in other models. Therefore, in the millimeter range the pulse profiles can be different from those at lower frequencies.

In parallel with spectral observations, it is necessary to measure the angular size of the emitting region in pulsars. Non-thermal pulsar radio emission is assumed to be generated inside the light cylinder with radius

\begin{equation*}
r_{\text{LC}}\left(\text{cm}\right)=\frac{\text{cP}}{\text{2$\pi $}}=4.8\cdot
10^9P\left(\text{s}\right),
\end{equation*}
\noindent where $P$ is the pulsar spin period.

Millimeter radio fluxes are low, smaller than 1 mJy. However, some
pulsars demonstrate outbursts with higher fluxes. Therefore, the
detection of a sufficiently bright outburst in the regime of
Earth-space interferometer would enable to resolve the light
cylinder region for the first time (up to pulsar distances of
several ten kiloparsecs) and would provide an invaluable information
for understanding the nature of these objects and localization of
the emission region of the observed electromagnetic radiation.

The angular size of the light cylinder is

\begin{equation*}
\theta \left(\text{arcsec}\right)=\frac{r_{\text{LC}}} d=3.2\cdot
10^{-7}\frac{P\left(\text{s}\right)}{d\left(\text{kpc}\right)},
\end{equation*}
\noindent where $d$ is the pulsar distance. With an angular resolution of 40
nano arcsec, the light cylinder of a pulsar with period 1 s can be
resolved from a distance of several kiloparsecs. The size of
emission region comparable with the light cylinder may additionally
support the synchrotron model.

\begin{figure}
\begin{center}
\includegraphics[width=7.5cm]{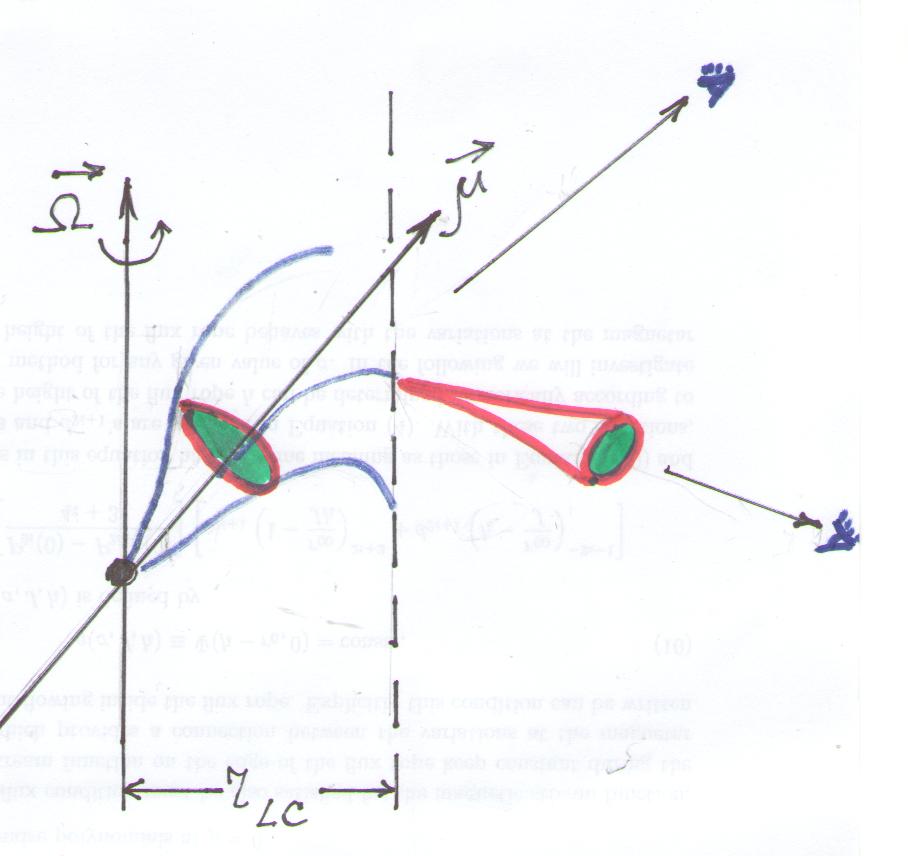}
\end{center}
\caption{Pulsar emitting only at millimeter wavelengths. The magnetic field lines are shown in blue. The dashed curve shows the light cylinder, the emission regions are shown in green. The upper observer sees the radio emission in the meter, decimeter and centimeter wavelengths, and the bottom observer will receive only the millimeter emission.}
\end{figure}

At magnetosphere periphery magnetic field lines are bended due to
plasma rotation, and it is possible to have a situation, where the
emission generated at moderate altitudes above the neutron star
surface cannot reach the observer, while the synchrotron radiation
at millimeter wavelengths can be seen by him. In that case the
appearance of a pulsar seen only at the millimeter wavelengths is
possible (Fig. 11).

The search for and detection of pulsars generated only microwave pulses is a totally new field in pulsar studies, which can significantly increase the observed neutron star populations.

Studies of pulsars with ground-space interferometer RadioAstron
suggest that there are interferometric responses exceeding the
diffraction spot size, i.e. when the scattering circle of pulsar
radio emission on the interstellar plasma inhomogeneities is
resolved. Here, the amplitude of the interferometer signal does not
decrease with increasing the interferometer base up to 240 000 km,
the maximum length obtained by the RadioAstron for radio pulsar
B0329+54 [82]. The structure of the interferometer signal with such
long bases is related to spectral parameters of the interstellar
plasma inhomogeneities. For distant pulsars, diffraction
scintillations can be observed at cm radio wavelengths. Observations
of such pulsars by Millimetron in the interferometric mode together
with the largest ground-based radio telescopes with 1 million km baselines
will probe both the structure of the interstellar plasma
inhomogeneities and that of the emission generation region in the
neutron star magnetosphere.

\subsection{Relativistic objects in centers of globular clusters}

Observations of central parts of the most massive globular clusters, in particular, with large core radii, in the single-mirror mode can be used to detect millimeter emission from stellar mass black holes in binary systems. Should the signal be detected, observations in the Earth-space interferometer mode with high angular resolution can be performed.

Black holes with masses (5-20) $M_\odot$ are end products of stars with
masses in the main sequence $/geq 25 M_\odot$ [83]. Recent theoretical
studies suggest that several hundreds of stellar-mass black holes
can be present in old globular clusters [84]. Almost all of them
should be single black holes, which is in agreement with a small
number of X-ray sources with black holes  in globular clusters. The
presence of black holes can heat up the central parts of the
clusters, leading to a significant increase in the cluster core
radius [85]. Therefore, old globular clusters with large core radii
are the most suitable to searches for stellar-mass black holes.

Recent observations with VLA (Very Large Array) discovered two black hole candidates in two globular clusters M22 [86] and M62 [87]. Both these clusters are massive but have different structure. For example, cluster M22 has an extended core, in line with theoretical predictions, while cluster M62 has a rather compact dense core.

VLA observations of M22 with a maximum possible angular resolution of $\sim 1 ''$  revealed that the observed faint source can be in a binary system (Fig. 12a) with properties similar to those hosting stellar-mass black holes (Fig. 12b).

\begin{figure}
\begin{center}
\includegraphics[width=7.5cm]{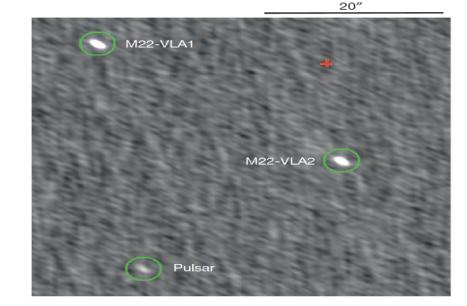}
\includegraphics[width=7.5cm]{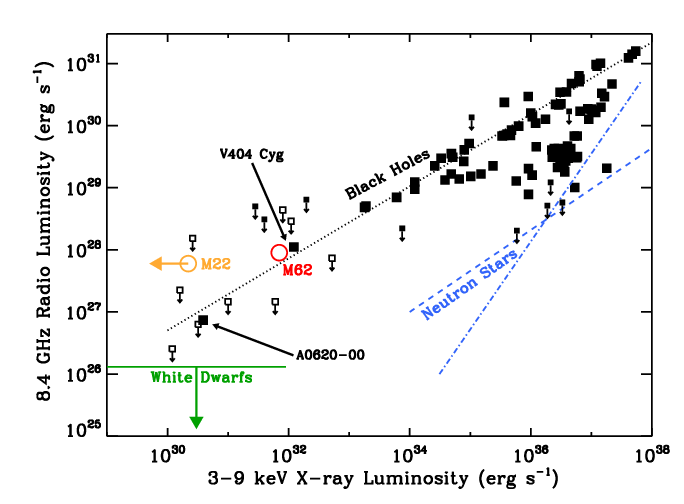}
\end{center}
\caption{(a) The VLA image of the globular cluster M22 in the continuum emission from the core. Two bright oval objects are sources identified as M22-VLA1 and M22-VLA2. The red cross shows the photometric center of the cluster [86].  (b) The X-ray to radio flux relation from stellar-mass black holes. $L_{8.4 GHz}^R$ is the radio luminosity at 8.4 GHz, $L_{3-9 keV}^X$ is the 3-9 keV X-ray luminosity.  The sources in M22 and  M62  have properties rather similar to black holes and not to white dwarfs or neutron stars [87].}
\end{figure}

VLA observations of the globular cluster M62 with a maximum time
exposure revealed the presence of a very faint central source
M62-VLA1 with a flat radio spectrum and radio flux 18.7 $\pm$ 1.9 $\mu$Jy at
the frequency 6.2 GHz [87]. Observed properties of this source in
the radio, X-rays and optical are similar to those of the well-known
transient X-ray source V404 Cyg, which is believed to be a
(presently quiescent) stellar-mass black hole in close binary
system.

High-sensitivity mm observations with a moderate angular resolution
of the black hole candidates discovered in globular clusters can be
extremely important to confirm their nature. Assuming a flat radio
spectrum, the expected mm fluxes from these sources can be a few
$\mu$Jy, while the fluxes from other stars should be much smaller. A
caveat here will be separation of the sources from the confusion
background  due to distant galaxies. Possible globular clusters to
be observed include M22, M14, M53, M62 and NGC2419.

\subsection{Origin of ultraluminous X-ray sources}

Search for intermediate-mass black holes can be another exciting task. Observations of possible mm and submillimeter  emission from ultraluminous X-ray sources (ULXs) and hyper-luminous X-ray sources (HXSs) in other galaxies in the single-mirror mode with possible subsequent high-resolution observations in the Earth-space interferometer mode should constrain the models and can help to unveil the nature of these sources.

Ultraluminous X-ray sources are off-center point-like objects with
an observed bolometric luminosity $L$ exceeding the Eddington limit
for galactic stellar-mass black holes ($20 M_\odot$), $3\times10^{39}$ erg s$^{-1}$, in the
0.3--10 keV energy range. ULXs were discovered in nearby galaxies by
the Einstein X-ray observatory [88]. For a spherical accretion of
fully ionized hydrogen, the Eddington limit can be expressed as
[89]:
\[
L_\mathrm{Edd}={4\pi c GMm_p \over \sigma_T \approx 1.3\cdot 10^{38} \left( {M\over M_\odot}\right) \mathrm{ergs s}^{-1}},
\]
\noindent where $\sigma_T$ is the Thomson scattering cross-section, $c$ is the speed of light, $G$ is the gravitational constant, $m_p$ is the proton mass, $M$ is the black hole mass. This implies that ULXs can harbor intermediate-mass black holes, $M = 10^2-10^4 M_{\odot}$ (see, for example, [90]). They also can be close binary systems  at an evolutionary stage that is not observed in the Milky Way. The number of ULXs is quite high: 230 [91]; in addition, more than 500 ULX candidates are discovered [92]. These objects are found in almost quarter of galaxies [93] of all types: in star-forming galaxies (which contain about 60\%  of all ULXs) [94]; in dwarf galaxies (Holmberg II); in elliptical galaxies with low star formation rate, as well as away from star formation regions (NGC 1313 X-2 and NGC 4595 X-10) [95].

Immediately after the discovery of ULXs, different models of their
origin were proposed. These models can be conventionally separated
in three main classes: subcritical accretion on to an
intermediate-mass black hole [96], supercritical accretion on to a
stellar-mass black hole [97] and collimated emission from a
stellar-mass black hole accreting at about the Eddington limit [98].
Since ULXs with different properties are observed, it is quite
possible that the ULX population includes different types of
objects. Stellar-mass black holes are the most likely candidates,
and intermediate-mass black holes can explain the observed
properties in several exclusive cases [99].

It is important to note that the existence of intermediate-mass
black holes is an unsolved issue in astrophysics. Until recently,
there has been no direct observational evidence of their existence,
as well as indirect indications of their reality, despite
predictions of the modern theory of structure formation in the
Universe. Thus, intermediate-mass black holes are a missing link
between stellar-mass black holes and supermassive black holes
located in galactic centers. According to current models,
intermediate-mass black holes could be formed both during collapses
of the primordial stars in the Universe and during the collapse of
dense cores of young stellar clusters [100], as well as a result of
accretion or stellar-mass black hole merging. In the last decades,
indications of the presence of intermediate-mass black holes in the
centers of globular clusters and in star-forming regions have been
emerged.

Unresolved central parts of massive globular clusters and HLSs are
the most probable intermediate-mass black hole candidates. In one of
the most probable intermediate-mass black hole, ESO 243-49 HLX-1,
discovered by XMM-Newton (X-ray Multi-Mirror Mission) space
observatory in 2004 in the spiral galaxy ESO 243-49 situated at 8
arcseconds from its nucleus [101], radio observations by ATCA
(Australia telescope Compact Array) discovered variable radio
emission at frequencies 5 and 9 GHz [102].

From some nebulae around ULXs radio emission was discovered. The most well-studied examples are Holmberg II X-1 [103] and NGC 5408 X-1 [104]. Both nebulae show optically thin synchrotron radio emission, similar to radio emission from supernova remnants. Different radio sources show different morphology, but in some cases they look like binary sources [95].

Millimeter observations of the nearby ULXs with a sensitivity of $\sim 1$ $\mu$Jy can provide more detailed information that is needed to estimate the black hole mass.

An unexpected result was recently reported by the NuSTAR (Nuclear
Spectroscopic Telescope Array) observatory [105]. One of the ULXs in
galaxy M82 turned out to be an X-ray pulsar ($P_{pulsar} = 1.37$ s) in a binary
system with orbital period 2.4 days. The mass function of the X-ray
pulsar is $f(m)\sim 2M$. Most likely, this is a neutron star with a high
magnetic field and strongly non-spherical magnetically collimated
accretion onto accretion columns near the neutron star surface.

\subsection{Supernovae}

One of the most plausible mechanisms of supernova
explosions is the magneto-rotational mechanism [106-108]. In this
mechanism, the explosion asymmetry arises due to the magnetic field
mirror symmetry breaking [109].  In first days after a supernova explosion its remnant is still very compact. Early observations of the supernova
remnant can be used to measure the initial explosion asymmetry
before an interaction of the remnant with the surrounding medium
that affects significantly particle trajectories of the expanding
remnant.  The Millimetron space observatory in the
interferometer mode will be capable of resolving the supernova
remnant up to a distance of 10 Mpc. The expected number of events is
several tens per year for type II supernovae and one type Ib/c
supernova per year.

Type IIn supernovae could also be interesting for Millimetron
observations. Such supernovae can be related to a common envelope
stage in the massive close binary evolution [110 -- 112].  The
common  envelope leads to asymmetric outflow of the external parts
of a red supergiant. The shock wave propagation along the expanding
red giant atmosphere can be observed in the millimeter wavelength
range. In this model, the outflow asymmetry can be detected starting
from the second day after the explosion for a supernova located at
distances above 30 Mpc.

\section{Black holes and jets}

Black holes are one of the most intriguing predictions of General
Relativity, and the problem of proving or refuting their existence
is a major task in astronomy. Black holes are extremely compact
objects, and to observe them a very high angular resolution is
needed. For all currently known objects the angular size of the
black hole horizon is less than 20 micro arcsec. The progress in
this field, hopefully, will bring the answer in the near future. For
example, the ground-space interferometer RadioAstron with a record
angular resolution of 7.5 micro arcsec is already carrying out
observations of the supermassive black hole in the center of galaxy
M87. The ground-based EHT, operating in the millimeter diapason,
will possibly resolve the central black hole in the Galaxy.

The space Millimetron observatory, operating jointly with ground-based telescopes, will enable measurements with an ultrahigh angular resolution and hence will be capable of probing smaller physical scales and a larger number of objects beyond these two nearby black holes. In addition to supermassive black holes, stellar-mass black holes in binary systems will be observed. These observations will help to probe the conditions of ultra-high energy particles generation and of the jet formation in the vicinity of black holes.

The VLBI specificity has been taken into account when planning the
Millimetron science. We assume that such an interferometer can be
used to estimate angular sizes of sources with ultrahigh angular
resolution at submillimeter  wavelengths, and with corresponding
sensitivity (see Table 1)  --- to construct maps in the frame of a
priori model assumptions of the source structure. The interferometer
can also be used to construct one-dimensional intensity maps.

Critically important information on the magnetic field and non-relativistic particle number density in the vicinity of the nearby black holes can be obtained not only in the VLBI mode, but also by polarization spectral measurements in the single-dish mode. The presence of a magnetic field in plasma along the line of sight results in the Faraday effect of rotation of the polarization plane of a linearly polarized light. This effect is qualitatively characterized by the Rotation Measure (RM), which is $\phi/\lambda^2$ where $\phi$ is the angle of the polarization plane turn, $\lambda$ is the wavelength.

According to theoretical studies [113, 114], the magnetic field on the scale of a black hole accretion disk can be $\geq 10^4$ G. Experimental papers [115, 116] showed that the magnetic field indeed increases towards the source nucleus. Therefore, if there are thermal electrons in the near nuclear region, extremely large vale of RM can be expected.

Under quite reasonable assumptions on the magnetic field value and the thermal electron density, it is easy to estimate that RM$>10^4-10^8$ rad m$^{-2}$.  The latest results confirm theoretical predictions of the extreme values of RM both from the Galactic center [117, 118[ and from distant active galaxies at millimeter wavelengths [119]. For linearly polarized synchrotron radiation from the nuclear region, narrow-band frequency channels of the Millimetron detectors will observe ``sine-like'' variations of the radiation flux density as a function of wavelength.

Detection of the extreme Faraday rotation by ground-based telescopes
or by Millimetron in the single-dish mode will provide a list of
candidate sources with definitely small angular sizes, which
subsequently can be observed by Millimetron in the interferometer
mode.

The origin of supermassive black holes in galactic nuclei is discussed in Sections 7.3 and 7.6.

\subsection{Nearby black holes}
Almost all nearby supermassive black holes (at distances less than 50 Mpc)  have a luminosity below the Eddington limit, which is related to a comparatively  weak accretion. These black holes include, in particular, the black hole in the Milky Way center (Sagittarius A*), M87, Centaurus A.

When modeling such sources the accretion rate is assumed to be many
orders of magnitude smaller than the Eddington value. For such small
accretion rates the standard disk accretion theory is inapplicable
and the so-called advection-dominated and radiative-inefficient
models are studied. In these models, the accretion disk near the
black hole is assumed to be geometrically thick with the
characteristic thickness of the order of the radial distance to the
black hole. As rarefied plasma of the  disk radiates ineffectively,
the matter heats up and expands in the vertical direction. The gas
temperature can be as high as $10^{11}-10^{12}$ K, and most of the released
accretion energy is advected by the radial inflow into the black
hole. Therefore, the efficiency of the released energy conversion
into radiation is very small, typically of order of $10^{-6}-10^{-3}$.

Since the accreting plasma is very rarefied, the optical depth for
free-free processes and Thomson scattering is believed to be small.
Moreover, in the submillimeter  range it is possible to neglect the
synchrotron self-absorption as well. This means that the medium near
the black hole in the submillimeter  range is optically thin, and
the black hole can be directly observed. Therefore, it becomes
possible, in principle, to determine the black hole mass, angular
momentum, as well as parameters of the accreting flow and its
geometrical structure. With sufficiently high angular resolution and
high sensitivity, in such flows it can be possible to directly
observe jet formation and matter outflows and, thus, to shed light
into so far unsolved issue whether the jet results from
magneto-hydrodynamic processes in the disk or arises due to the
so-called Blandford-Znajek effect related to the black hole
rotation. Provided that scales smaller than the gravitational radius
can be probed, a principal possibility of studying turbulence of the
accretion flow and understanding of the related phenomenon of
quasi-periodic luminosity oscillations arises, which can be due to,
for example, the presence of hot spots in the disk and/or the
excitation of different oscillation modes in the accretion flow by
turbulence.

The best studied object of type is Sgr A* (Fig. 13) in the Galaxy.
The distance to this black hole is $R=8$ kpc, its mass is $4\times10^6 M_\odot$, the
gravitational radius is $r_g\sim10^{12}$ cm, the bolometric luminosity is $3\times10^{36}$ erg
s$^{-1}$, the angular size of the horizon is $r_g/R\sim 10$ micro arcsec. The
variability time-scale varies from several minutes to several hours
in the near infrared, X-ray, submillimeter  and mm ranges.

\begin{figure}
\begin{center}
\includegraphics[width=7.5cm]{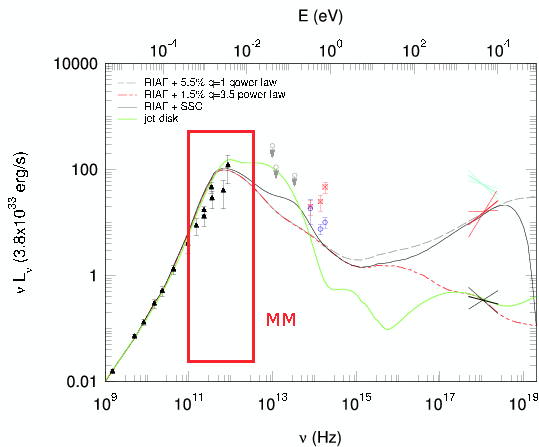}
\end{center}
\caption{The spectrum of the Galactic center (http://www.mpe.mpg.de/368843/Results). Radiation inefficient accretion flow (RIAF) with the fraction (in per cents) of non-thermal electrons with power-law spectrum characterized by the slope $q$. SSC (Synchrotron Self-Compton) model takes into account Compton scattering of synchrotron photons. The symbols show radio and IR-measurements, the crossing segments show the X-ray spectral measurements. The red quadrangle shows the Millimetron range.}
\end{figure}

There are several breakthrough tasks for Millimetron in the
astrophysics of supermassive black holes. The primary task is to
resolve the gravitational radius almost for all supermassive black
holes within 50 Mpc. The Millimetron space observatory will be
capable of resolving, in principle, a structure of accreting flows
around black holes at scales of the order of gravitational radius
for about 40 supermassive black holes. For such objects as the
supermassive black holes in the Galactic center or in M87, the
angular resolution of Millimetron will be sufficient to study
details a few hundred times smaller than the gravitational radius
(Fig. 14). This enables us to probe turbulence in the surrounding
gas flows, which is a major issue in the theory of accretion.

\begin{figure}
\begin{center}
\includegraphics[width=7.5cm]{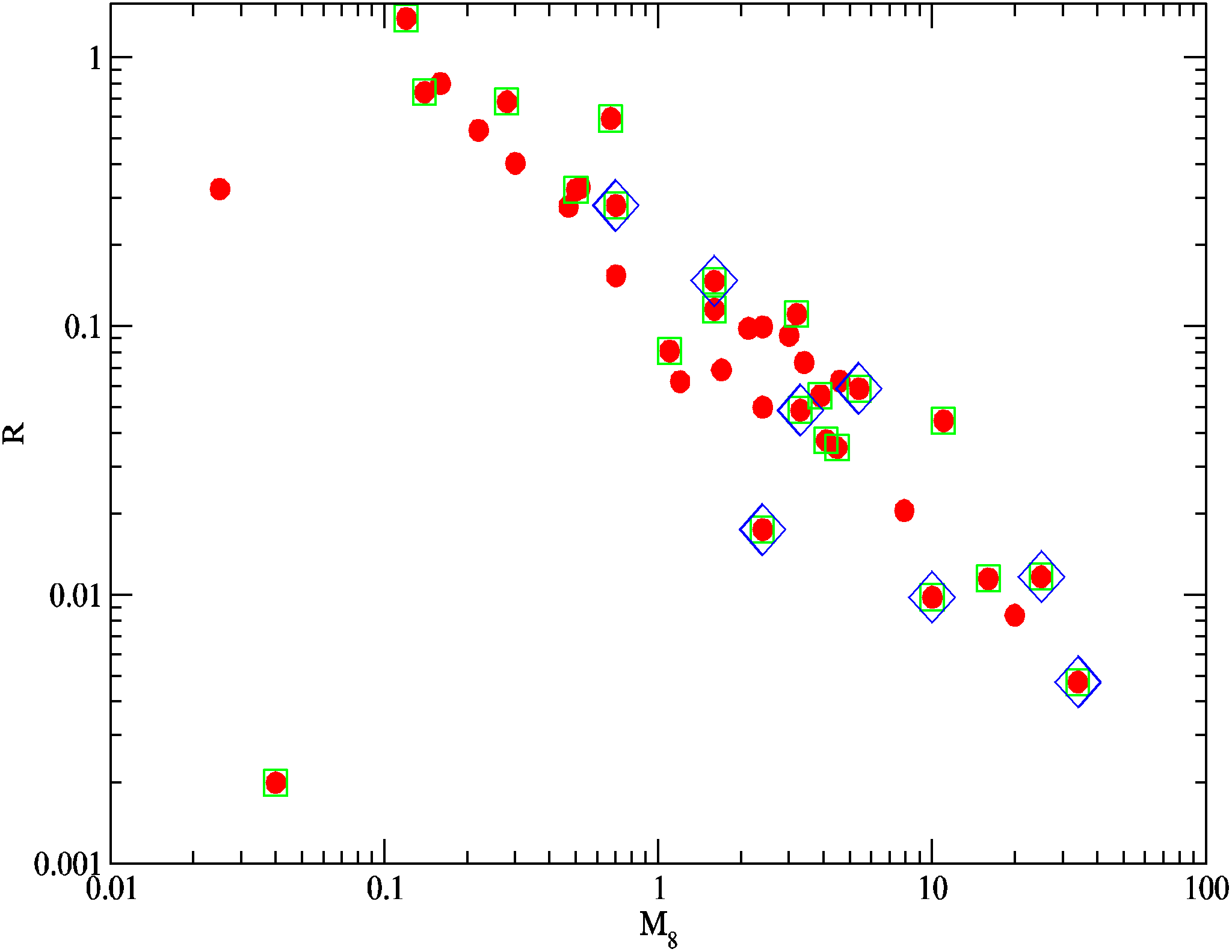}
\end{center}
\caption{The ratio of the minimum angular size that can be resolved by Millimetron to the gravitational radius for 38 supermassive black holes at distances within 50 Mpc. The minimum angular size is assumed to be $2\times10^{-3}$ rad. The filled circles correspond to the sensitivity threshold in the interferometer mode $10^{-4}$ Jy, filled squares correspond to $10^{-2}$ Jy, diamonds correspond to $10^{-1}$ Jy. The square in the left bottom corresponds to the central black hole in the Galaxy, the bottom right diamond shows the black hole in M87 [120].}
\end{figure}

\subsection{Black holes shadows}

As noted in Section 5.1, if gas near a black hole is optically thin,
the black hole can be seen directly. Indeed, some radiation of this
gas is captured by the black hole, and so the region around the
black hole at large distances should appear as a dark spot
corresponding to the captured radiation. The brightness distribution
around this spot will be very strongly changed due to a strong
distortion of the light trajectories near the black hole,
gravitational redshift, etc. The brightness distribution around the
dark spot can be used to study main characteristics of the emitting
gas, for example, to determine whether disk or jet mainly
contributes to the object luminosity at a given wavelength. A form
of the brightness distribution around the dark spot is also strongly
dependent on the black hole spin, and a study of this region is
apparently the most direct tool of determination of this principal
parameter. However, the most significant provided result is that the
discovery of the black hole shadow will be the direct evidence that
superdense and supermassive objects in the centers of the galaxies
are indeed black holes. The angular resolution of ground-based
interferometric systems is limited by the Earth diameter, and
generally is sufficient to search only for shadows around the
closest two-three supermassive black holes, whereas observations
from Millimetron in the interferometer mode offer the principal
possibility to discover shadows around several dozens of such
objects (see Fig. 14).

Fig. 15 shows the expected shapes of shadows from the central black
holes in our Galaxy and M87, which were calculated numerically. In
Fig. 15a, the black hole illumination is due to a radiatively
ineffective disk, and in  Fig. 15b --- due to a jet. Calculations
were carried out taking into account parameters of the EHT
telescope. It is very important to perform similar calculations for
parameters of the Millimetron space observatory.

\begin{figure}
\begin{center}
\includegraphics[width=8.5cm]{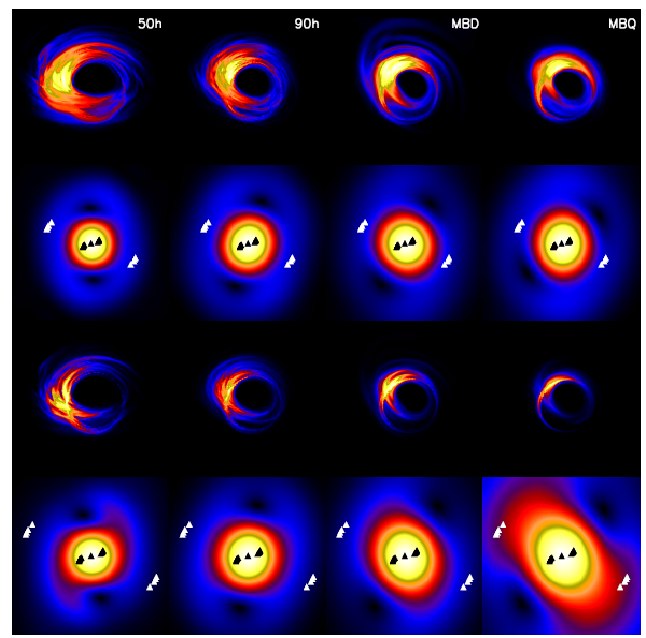}
\includegraphics[width=6.5cm]{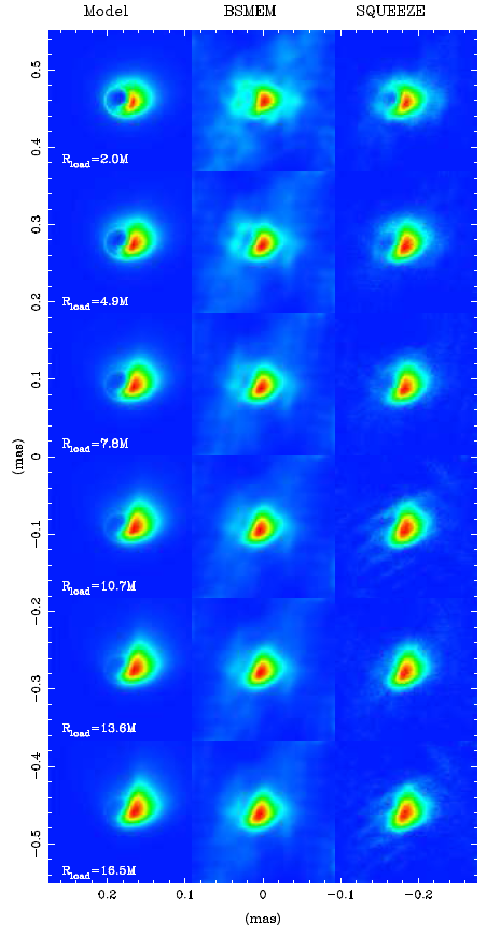}
\end{center}
\caption{(a) Model image of a radiation-ineffective accretion disk and shadow from the central Galactic black hole. Shown are the intensity distribution (1st and 3d rows) and the visibility function amplitudes (2d and 4th rows) for different models of the accretion flow (in different columns). The models differ by the black hole spin parameter (from left to right): 0.50, 0.90, 0.92, 0.94. The upper and bottom two rows are calculated for wavelength 1.3 mm and 0.87 mm, respectively [121]. (b) The brightness distribution of the central region of galaxy M87.  The black hole  is assumed to be ``illuminated'' by jet. Calculations for the frequency 345 GHz. The left column shows the initial model distributions, the central and right columns show  the model interferometric images as obtained by different methods. $M$ is the black hole mass. Geometrical units with $G=c=1$ are used.  The physical units are obtained by multiplying the mass by $G/c^2$. The jet formation radius from the black hole decreases from bottom to up [122].}
\end{figure}

\subsection{Distant black holes}
Millimetron will offer the possibility to resolve gravitational
radius of a black hole with mass of 100 million solar masses located at
a distance of  100 Mpc, and of a black hole with mass of 1 billion solar
masses from a distance of 1000 Mpc. A lot of bright active galactic
nuclei and quasars can be found at such distances, which presumably
contain black holes in this mass range, where accretion proceeds
through the standard geometrically ``thin'' accretion disk. The
presence of such a disk can be inferred from a ``Big Blue Bump'' in
the UV spectrum. Although these disks are optically thick, the black
hole shadows of  can also be searched for (Fig. 16), since the size
of such disks in the direction perpendicular to the disk plane is
quite small compared to the horizon angular size.

At large distances it is also possible to observe very interesting
non-stationary phenomena in active galactic nuclei and quasars,
which are not observed in relatively close objects due to their low
probability/ Let us briefly discuss only two such phenomena:
supermassive binary back holes (SBBHs) and tidal disruption events
of stars by a supermassive black hole.

SBBHs are not only interesting objects by themselves, but also can
generate the most powerful bursts of gravitational radiation during
their coalescence, which can in principle be detected by future
space based gravitational wave interferometers from very large
distances (up to cosmological scales).  Apparently, the most known
SBBHs candidate is the BL Lac object OJ 287 located at a distance of
1 Gpc. OJ 287 is thought to be a SBBH with component masses 10 bln
and 100 million Solar masses and an orbital period of 12 years (Fig.
17). In the case of this object Millimetron can easily resolve
scales of the order of the angular size of the horizon of the more
massive component.

\begin{figure}
\begin{center}
\includegraphics[width=7.5cm]{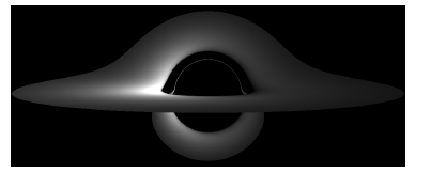}
\end{center}
\caption{The bolometric flux distribution from accretion disk around a Schwarzschild black hole. The disk inclination angle to the line of sight is 84.5$^\circ$ [123].}
\end{figure}

Tidal disruptions of stars by q black hole are usually manifested as
X-ray outbursts in inactive galactic nuclei. A characteristic decay
time of the outbursts is about a few years.

If the submillimeter  luminosity of these objects is sufficient to
be observed by Millimetron in the interferometer mode, estimates
show that Millimetron will be capable of resolving scales of order
of the gravitational radius for at least several objects, including,
for example, NGC5905 [124]. A very interesting source Swift
J1644+57, discovered in 2011 [125], is a very powerful source in all
wavebands, including the millimeter range, locating at a distance of
about 1 Gpc.  This source is interpreted as a jet formed after the
tidal disruption of a star by a black hole. Millimetron will be able
to test the origin of such events.

\begin{figure}
\begin{center}
\includegraphics[width=7.5cm]{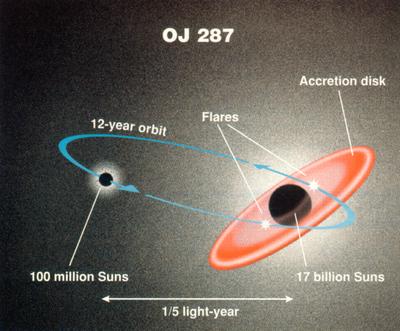}
\end{center}
\caption{Possible schematics of OJ 287 (http://www.astro.utu.fi/news/080419.shtml).}
\end{figure}

\subsection{Physics of jets}
Variety of accretion theories leads to variety of theories of jets. AS  in the case of disks, the Millimetron space observatory, thanks to its record parameters, will provide missing observational data needed to test different models.

In the case of accretion disks around black holes, the main issues
can be listed as follows (see, for example, [126]). At first, it is
necessary to understand the jet formation mechanism --- is it either
due to the black hole rotation or due to processes in the accretion
disk? Secondly, to understand origin of jet, a magnetic field
structure near the black hole should be known. Thirdly, it is yet
unclear what kind of plasma moves in the jet (is it
electron-positron or electron-ion plasma?), what is its
characteristic velocity as a function of distance to the black hole
and to the jet axis, what is a distribution of non-thermal particles
in the jet, etc. Fourthly, it should be reliably determined whether
``typical'' jets are two-side or one-side (i.e. whether the
formation of a jet directed only to one side is possible) [127].
Fifthly, it is important to obtain observational confirmation of a
possible change of the jet magnetic field polarity [128, 129]. All
these problems are interrelated.

In different numerical models, the very possibility of the jet
formation depends on the structure of accretion disk and of the
magnetic field (Fig. 18). Clearly, to unveil the nature of jets
observations with high angular resolution are required, which can
probe matter outflows and the magnetic field structure in the
vicinity of black holes. Note that when polarization is taken into
account the VLBI-mapping can be hampered by extreme Faraday
rotation. Possible methods of the mappings are proposed in papers
[131-134].

Jet formation studies are closely related to accretion disk studies,
and observations of these objects in different types of
astrophysical objects from stellar-mass black holes to quasars are
crucial. With record high angular resolution in the interferometer
mode, Millimetron possibly can solve these problems. We stress again
that in order to study the magnetic field structure, polarization
measurements are crucial.

\begin{figure}
\begin{center}
\includegraphics[width=7.5cm]{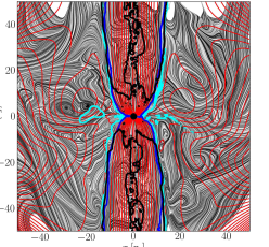}
\end{center}
\caption{Velocity and magnetic field distribution (shown in grey and red, respectively) according to the relativistic jet formation model near a black hole with spin parameter $a=0.9375$ [130]. On axes: coordinates in the black hole Schwarzschild radii.}
\end{figure}

Finally, millimeter observations can be decisive to understand the
nature of GeV and TeV-flares from active galactic nuclei [135]. Such
flares are observed from galaxies M87 and 3C454.3 each 2-3 years and
last from several days to several weeks. Ultrahigh angular
resolution observations can both localize the flare in the jet and
to follow the evolution of the emitting region. To clarify the
origin of the flares, it is also important to observe simultaneously
with such flares central regions of active galactic nuclei [127,
136-138].

\subsection{Jets from cosmic gamma-ray bursts}
An unprecedentedly high angular resolution of Millimetron in the
interferometer mode offers the possibility to directly observe jets
from cosmic gamma-ray bursts during first several days after the
burst, when the jet expansion is relativistic. Indeed, for nearby
GRB 030329, which is located at the redshift $z=0.1685$, the angular
resolution of 0.16 micro arcsec corresponds to 0.0005 pc, which is
equivalent to a light-time of 1 day. Thus, it becomes possible to
observe and to investigate the early phase of jet expansion,
and may be even before the so-called jet-break time when the GRB
afterglow power-low light curve does exhibit a break.

These observations will substantially complete the ground-based VLBI
observations [139] and will allow the determination of several
important parameters, such as the jet-break time, the time of the
jet deceleration, the time of the transition from relativistic to
the Newtonian expansion, as well as to test the possibility of
occurrence of two successive jets with different parameters [140-142]. 
Despite nearby gamma-ray bursts are quite rare, once per 5-7 years, it
might be possible to detect one such an event during the Millimetron
mission.

\section{Galaxies}
\subsection{Evolution of galaxies}
Star formation leads to dust production, therefore star-forming
galaxies are bright sources in the submillimeter range. On the other
hand, even a warm dust ($T_d=30$ K)  in early galaxies at redshifts $z\sim10$
will be seen in the submillimeter range. {The spectral line 158 $\mu$m of CII ion, 
which provides cooling of interstellar gas with temperatures 30 K to several thousands K
will also fall into the millimeter wavelengths.} This {makes} Millimetron with its high sensitivity a 
{powerful instrument for detection of} galaxies at high redshifts up to $z\sim 6-7$ and to {draw} a
sufficiently complete picture of their evolution. To achieve this
aim, both continuum and spectral line (CO, CII, OI, etc.)
observations are required. {Angular} resolution plays a key role {as well}.

Galaxies with active star formation are so numerous that under
insufficient angular resolution their images can merge.  For the Millimetron 10-m primary mirror this effect is
much weaker than for the {Herschel} telescope and the SPICA project
with 3.5-m mirrors (Fig. 1b). Preliminary estimates show that
Millimetron will be able to measure {in total} spectra of at least 10000
galaxies and to make continuum observations for several {tens of millions of} 
galaxies. Thus, it will {be able to} obtain three orders of magnitude more
information than the {Herschel} telescope. Millimetron will be able to
{detect light from} galaxies at redhsifts up to about $z=6-7$ (Fig. 19).

\begin{figure}
\begin{center}
\includegraphics[width=7.5cm]{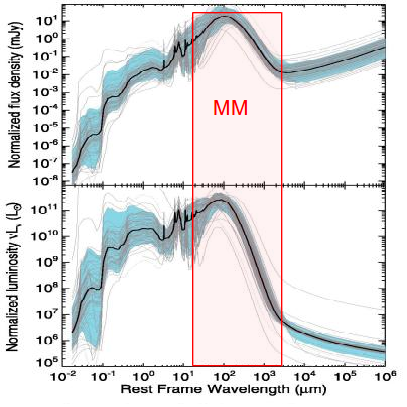}
\includegraphics[width=7.5cm]{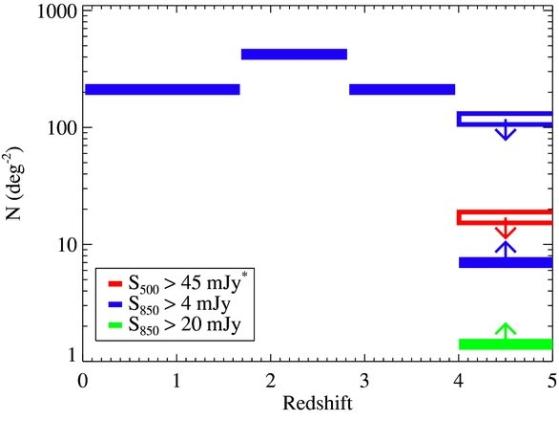}
\end{center}
\caption{(a) Spectral energy distribution from typical submillimeter galaxies (in the rest frame) peaked at a wavelength of $\lambda \approx$ 100 $\mu$m  [143]. The red rectangle shows the Millimetron operation range in the single-dish mode. (b) The distribution of submillimeter galaxies number density as a function of redshift [144]. $S_{500}$ and $S_{850}$ are the fluxes at 500 and 850 $\mu$m , respectively.}
\end{figure}

Lyman-alpha emitting galaxies represent another {class of} very interesting {objects representing violent phases of} galactic evolution. The source of emission in these galaxies and their place in  general scheme of galaxy evolution are still unknown. The issues to be solved include:

\begin{itemize}
\item what is the dust content in this galaxies?
\item what evolutionary {stages} do they {re}present?
\item what are the energy sources{: accretion on to a black hole, 
stellar emission, or gravitational energy?}
\end{itemize}

To answer these questions {millimeter and sub-millimeter} photometric and spectral observations of
 {such} objects, {identified primarily in optics are needed.} 
Due to high sensitivity Millimetron {detect and} study typical
objects with fluxes of the order of several tens of $\mu$Jy [145].

\subsection{Low star-formation regions}
{An outstandingly peak}  sensitivity of Millimetron enables measurements of
temperature and mass of dust along the line of sight in  {dilute galactic and intergalactic environment}, where dust is found
under specific conditions of a low-density {diffuse}  medium ({one to two orders} of magnitude {less dense} than near the Sun). It is important to
understand mechanisms responsible for presence of  dust {in such environment} 
and its heating. Studies of {dust in } galactic disk {outskirts}, in elliptical galaxies
and in galaxy clusters, where the dust formation process is
{inhibited}, will allow to clarify processes of dust and gas
transport into circumgalactic and intergalactic space, as well as to
understand {particular} features of gas molecularization and star formation
{in a} low-density interstellar matter.

Recent studies have revealed two types of {spatial} dust distribution at
galactic peripheries. In one case, the dust-to-gas mass ratio
decreases with radius proportionally to the metallicity, as in galaxies M99/M100 [146] or M31 [147, 148].  In {the} other
case,  the
dust-to-gas mass ratio remains {nearly} constant at distances up to
one and a half optical radius {even though the underlying metallicity goes down outwards} [149, 150]. The first case corresponds
to {\it in situ}, {i.e., along with metals}, dust production.  The second {though} 
 corresponds {either} to a selective dust transport {in the 
radial direction along the disk},  or an overestimate of its relative {mass} content. {If confirmed this} 
dichotomy in the spatial dust distribution at the galactic periphery {seems to} 
reflect {particular}  features of dynamical processes in
galactic disks, and can be of primary importance for {their} 
evolution.

\begin{figure}
\begin{center}
\includegraphics[width=7.5cm]{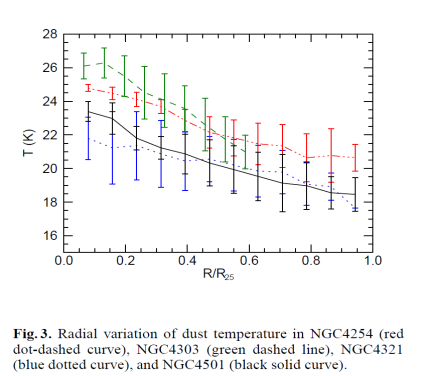}
\end{center}
\caption{Radial variation of the dust temperature in four galaxies: NGC 4254 (red curve), NGC 4303 (green curve), NGC 4321 (blue curve) and NGC 4501 (black curve) [151]; $R_{25}$ is {radius to the 25$^{\rm th}$ magnitude isophote}.}
\end{figure}

IR-radiation fluxes from dust at the periphery  suggest 
dust temperature of $\lesssim 20$ K ( Fig. 20). {IR fluxes from dust} 
beyond the galactic disk radius, {commonly assumed} to be its photometric
radius, are {as a rule}  below the {Herschel} sensitivity, however it
{seems to} be {quite} sufficient to be detected by Millimetron. Addressing this
task by Millimetron will allow:

\begin{itemize}
\item to find the source of  dust heating in the ISM as a whole, {and} at the periphery beyond stellar disks;
\item to study optical properties of dust at the galactic periphery, which will help to understand the dust transport ({and/}or production) mechanisms far from the main dust production sources;
\item to study the {gas-to-dust} {ratio} as a function of the local star formation rate, the surface gas density and its  {total amount}. This in turn will give a new  {approach} to estimate the total mass of molecular gas and to understand its role in the star formation process;
\item to evaluate the mass of {CO-dark} molecular gas {which is not traced by CO emission}, {though manifested by gamma-ray and ionized carbon emission, and by presence of dust in there} 
 [23, 152]. {Such} low-density {CO-dark} regions can
contain most of the molecular gas. The example of the Galaxy
shows that fraction of CO-dark molecular gas increases with
decreasing gas density, reaching 80\%  at a distance of 10 kpc from
the center [23];
\item  {with making use the dust-to-gas ratio} to determine  density and spatial distribution of {the} ISM and to connect {them} 
to the observed star formation rate in low-density gas regions: at {outlying} 
periphery of galactic disks [153], in tidal structures (tails
and bars of interacting galaxies) and in the vicinity of interacting
galaxies in the intergalactic space where star-forming regions {are}  
also  found [154]. Of a special interest are studies of faint
dust emission in very low-surface brightness spiral galaxies ({such as} 
Malin-1, Malin-2) and  with very low column density of HI
in disk which, nevertheless, have  spiral structure and at
least in some cases contain molecular gas detectable in CO-lines
[155]. Faint FIR emission has been registered only from a few such
objects [156]. The dust mass estimate will  {make possible to determine CO-dark} molecular gas
content and its {spatial} distribution in
these objects, which is important to explain spiral structure
and  low star formation rate in {their} disks, as well
as details of their evolution [157];
\item to carry out a detailed investigation  of FIR emission from star formation regions {nearby} 
interacting galaxies (in bridges and connecting bars as observed,
for example, around the Antennae galaxies or in the M81/M82 group),
where  {much} dust can be found. {Such study seems to be of principal importance as the dust } can affect gas
thermodynamics {and stimulate}  a star formation burst. In {the tidal bridge of M81/M82 group}
 {dust has been observed in extinction} with
the dust-to-gas ratio six times {of} the standard value in the
Galaxy [158]. Unfortunately, this region of the sky  {might be contaminated by effects from }  Galactic {cirrus clouds} (clouds of {relatively} low
surface density above the Galactic plane), which complicates a
correct interpretation of the extinction measurements. Since  gas
and dust temperature in {this tidal bridge  seems to be}  low, the Millimetron observations
could give a definite answer on the amount of dust in the M81/M82
region.
\end{itemize}

{Study of general properties of dust: its mass, optical characteristics, spatial distribution --- for low-surface brightness galaxies are of great importance for understanding physics of galaxies in general. This 
not only will provide estimates of the amount of CO-dark molecular gas, which  is essential for 
correct dynamical modeling of such galaxies and their dark matter halos, but bring also  
understanding star formation mechanisms under the conditions 
when standard criteria are not met. IR-fluxes
expected from low-surface brightness galaxies are below the
Herschel sensitivity limits, but seem to be measurable by
Millimetron.} 

\begin{figure}
\begin{center}
\includegraphics[width=7.5cm]{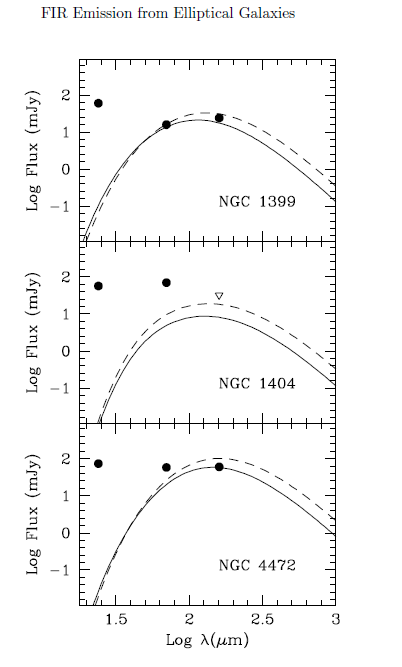}
\end{center}
\caption{Comparison of fluxes $F$ at 24, 70 and 160 $\mu$m  (filled circles) for three elliptical galaxies with model spectral energy distribution [159]. Dust components with different temperatures are clearly visible.}
\end{figure}

Elliptical galaxies with low dust content and almost without cold
gas provide another interesting {field} for  studies {with Millimetron}. Current
data suggest that dust is present there {in the form of} {and includes several populations} with different temperatures (Fig. 21). Due to a low gas
density, dust in elliptical galaxies immediately {enters} into  the
hot gas phase and turns out to be {unshielded of} destructive effects {from} 
the hot gas. From this point of view, elliptical galaxies represent
a unique laboratory, where the dust destruction mechanisms must {be} 
clearly {revealed}. On the other hand, in a low-density {environment} {of} ellipticals, where collisional friction is weak 
dust transport by radiation pressure is clearly manifested,
{study of spatial} dust structure in {the} galaxies
will {allow} to understand dust redistribution and its total ``budget''
in the Universe.

{Study} of dust  in {the Universe} have another very important aspect:
dust is responsible for optical absorption, and for correct
interpretation of many optical observations dust distribution should
be well known. For example, this is of principal importance for
cosmological supernovae projects aimed to determine 
the {dark energy parameters} [160, 161].
Different estimates {of dust content} are controversial {by factor of several} 
[162, 163]. For example, {the} analysis of about $10^4$ galaxy clusters
from SDSS (Sloan Digital Sky Survey) at low redshifts  (0.1-0.2)
with all background quasars within 1 Mpc around the cluster center
[164] suggests the mean extinction $A_v=0.003\pm0.01$. Close result is obtained
from the analysis of 90000 {SDSS background} galaxies {and 
458 foreground} galaxy clusters at redshifts up to $z\sim0.5$ [165]. The corresponding
mean mass dust fraction  is $\rho$(dust)/$\rho$(baryons) $\sim10^{-5}-10^{-4}$, i.e. 0.1-1\%  of
the Milky Way value. In this connection, an  interesting (and
intriguing) {circumstance} is the measured dust extinction in the intergalactic
space: $0.03<A_v<0.1$ [166], i.e. an order of magnitude as high as in the {intracluster gas}. This {result}, if true, suggests that either dust
is supplied into the intergalactic space by field galaxies (i.e.
{those} that do not belong to clusters or groups), or a significant
fraction of dust in galaxy clusters is unobservable (being
apparently bound in dense cold clouds with small geometrical
cross-section). Both possibilities are worth of a careful
investigation.

On the other hand, from the analysis of about 7000 clusters
and groups from SDSS at low redshifts ($z=$0---0.2){ [167]}, the dust mass
fraction has been derived: $\rho$(dust)/$\rho$(baryons) from 5\%  to 55\%, 
{of} the local (Galactic) value {
for clusters and groups, respectively}. {Even though} for galaxy groups this high value {looks quite reasonable}, since the gas temperature in {them}  is below the
critical limit of {efficient} dust destruction, the higher dust
concentration in clusters measured in [164,165] {might} suggest that dust in clusters is shielded from
destruction by dense cold {gaseous envelopes} of cloud fragments {along with which dust 
has been expelled from galaxies}. 

\begin{figure}
\begin{center}
\includegraphics[width=9.5cm]{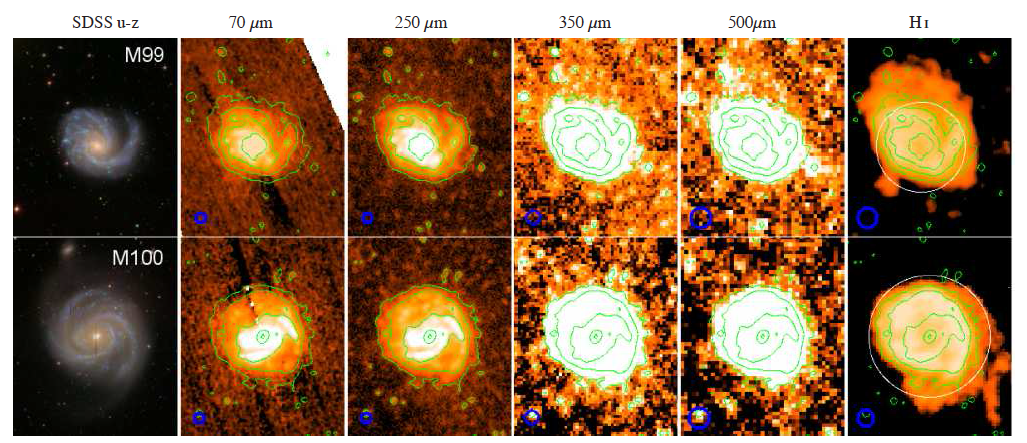}\\
\includegraphics[width=9.5cm]{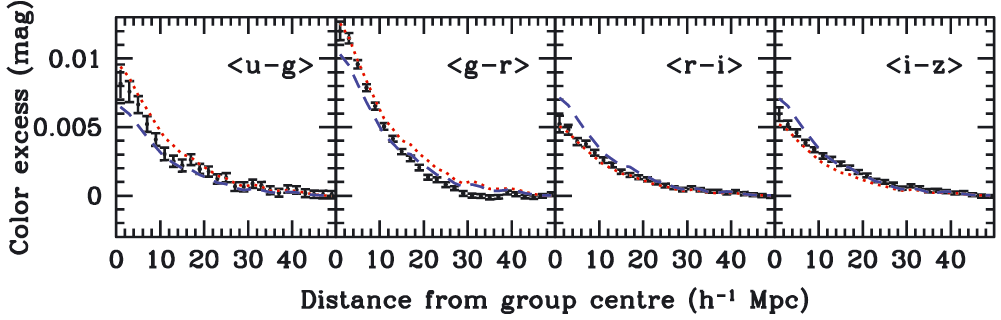}
\end{center}
\caption{(a) Correlation of the intensity of PAH emission with metallicity according to observations of extragalactic star formation regions. The PAH emission is represented by the parameter $P_{8,0}$, the flux ratio at 8 $\mu$m  to the total flux above  70 $\mu$m . To determine $P_{8,0}$, FIR observations with the same angular resolution as in near IR are needed [174]. (b) Image of galaxy M82 in the line of ionized carbon at 158 mm in celestial coordinates $\alpha,\;\delta$ (epoch J2000) [175].}
\end{figure}

The situation {does not become} completely clear even when we {turn} to
IR-observations in emission, although some {indications} to a possible
{solution do appear}: Spitzer space telescope data at wavelengths $\lambda=$24
and 160 $\mu$m from the Coma cluster direction do not {unexpectedly} show {signal} above the noise level [168]. However, later
observations by  {Herschel} observatory at wavelengths $\lambda=$ 100, 160, 250, 350, 500 $\mu$m
revealed dust traces in the vicinity of several dwarf galaxies
with extended halos in the FIR band {through clear demonstration} of 
{flattening} the gradient at long wavelengths. The extended FIR
emission with ``colder'' spectrum at the periphery of elliptical
galaxy M87  interpreted {initially} in terms of cold
dust emission, later was identified with synchrotron radiation
[169]. The example of the M87 galaxy shows that {FIR} and
millimeter spectra of extended disks of spiral and elliptical
galaxies can {stem from} a superposition of cold thermal dust
emission and non-thermal emission {of} relativistic electrons {spread 
diffusively in disk outskirts.}  From this point of
view, submillimeter and millimeter observations could be crucial,
since the difference between {contributions of} Jeans thermal dust
spectrum and power-law spectrum of relativistic electrons with
negative slope is most pronounced { in these bands}. In this connection, the latest
results of cold dust observations in coronae and extended disks of
nearby isolated galaxies M31 [147, 148], M99/M100 (Fig. 22a) [146],
in dwarf galaxies from the Virgo cluster [169, 170] should be
mentioned. Similar dust coronae around galaxies are well observed in
optics [167].

{Recently,} numerous evidences of the
{presence} of extended (up to 300 pc in
{radius}) {low-redshift (0.1-0.4)} circumgalactic {gaseous} coronae enriched with heavy elements up to
almost the solar metallicity {have been found  [171, 172]}. The dust should be expected
to be present in {such} coronae in the same proportion as
metallicity. Taking into account that the gas temperature in the
coronae is not very high, typically around $10^6$ K [173], a significant
dust fraction can be {survived against destruction}. Therefore, dust observations in such
coronae could provide valuable information on the dust evolution
during its transport into the intergalactic medium.

\subsection{Extragalactic star formation regions}
Observations of star formation regions in dwarf galaxies similar to Holmberg II and DDO 053 require a higher angular resolution than available with the {Herschel} space telescope --- at least as high as 10-20 arcsec at wavelengths longer than 200 microns. To observe low-metallicity star formation regions, the sensitivity level should be better than 1 $\mu$Jy per bin. Such Millimetron observations will be unprecedented and can bring breakthrough results.

A high angular resolution in the near-IR range {has been} attained by the
Spitzer telescope. However, in order to fully describe dust {properties} 
similar angular resolution {in  FIR} is needed {as well}. Observations of polycyclic
aromatic hydrocarbons (PAH), in particular, study of their
abundance dependence on the ISM parameters in galaxies can be a
perspective task. {Such} observations will clarify reliability of the
PAH emission as star formation rate indicators, and address the
PAH survival and organic species evolution in ISM {as well}.

In individual galaxies and star formation regions the PAH abundance
correlates with metallicity (Fig. 23a), however the nature of this
correlation is unclear. The dependence of the PAH formation
processes on metallicity in stars or molecular clouds, and processes
of their destruction in ISM and star formation regions are
considered as possible explanations. The problem is significantly
complicated {though} due to the lack of high-resolution observations (at
least 10 arcsec) at wavelengths above 200 microns. The analysis of
the PAH content, i.e. their contribution to the total mass of the
dust, is impossible without these data.

\begin{figure}
\begin{center}
\includegraphics[width=7.5cm]{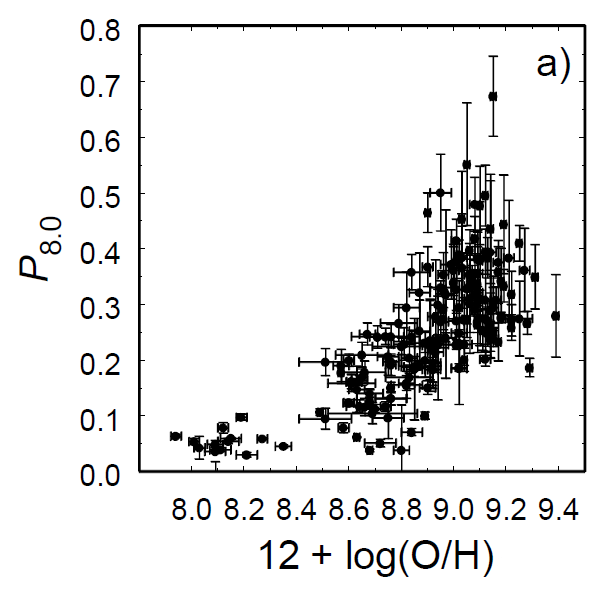}
\includegraphics[width=7.5cm]{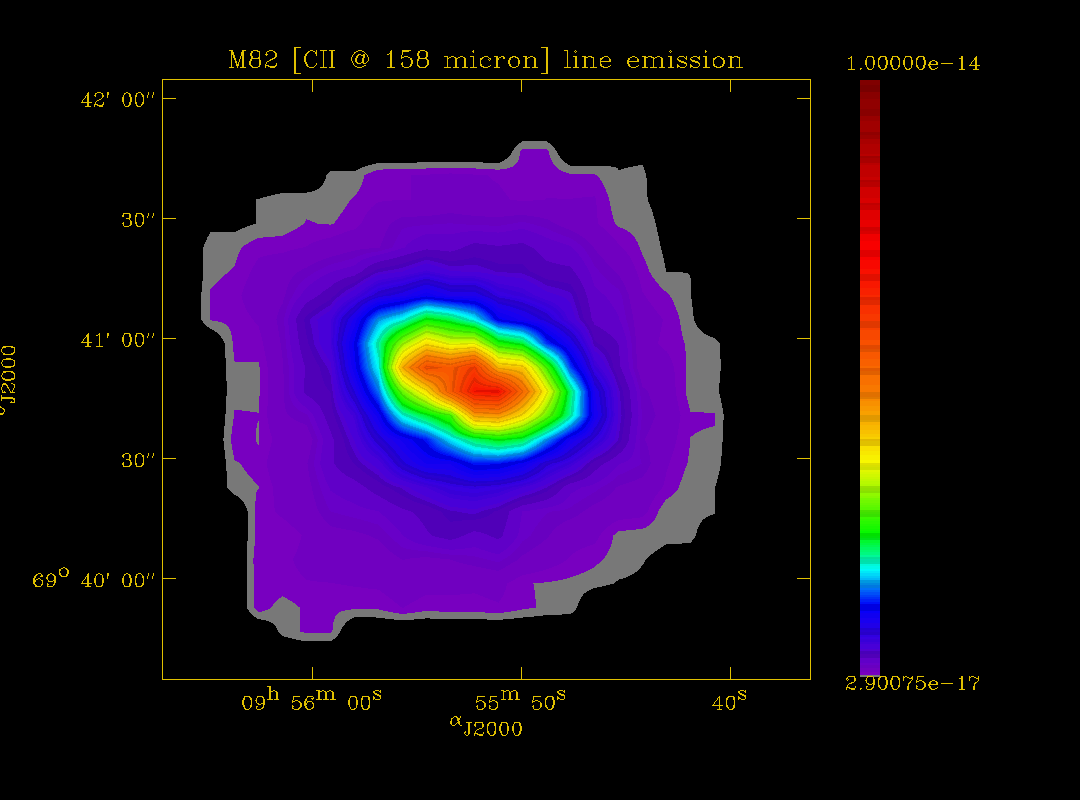}
\end{center}
\caption{(a) Correlation of the intensity of PAH emission with metallicity according to observations of extragalactic star formation regions. The PAH emission is represented by the parameter $P_{8,0}$, the flux ratio at 8 $\mu$m  to the total flux above  70 $\mu$m . To determine $P_{8,0}$, FIR observations with the same angular resolution as in near IR are needed [174]. (b) Image of galaxy M82 in the line of ionized carbon at 158 mm in celestial coordinates $\alpha,\;\delta$ (epoch J2000) [175].}
\end{figure}

For extragalactic star formation regions the most important lines
are [CII] (158 $\mu$m), HD (112 $\mu$m), HeH$^+$ (149 $\mu$m, 74
$\mu$m) , H$_3^+$ (95 $\mu$m ), H$_2$D$^+$ (219 $\mu$m, 207 $\mu$m). For example,
Fig. 23b shows an image of the star-burst galaxy M82 in the ionized
carbon line at 158 $\mu$m obtained by the {Herschel} space telescope.
It is seen that low angular resolution enables studies {only} of a 
general distribution of emission {over the} galaxy but not of
individual star-forming regions.

Observations by {Herschel} telescope and by other millimeter
observatories have revealed the {existence of} dust even there
where its presence has not been {expected} before: in elliptical
galaxies, where dust must have been destroyed and at the
periphery of disk galaxies far from  star formation regions.
New {highly sensitive} measurements of weak dust emission at
galactic periphery, as well as observations of protostellar and
protoplanetary objects are required to clarify these issues.  The
main method of research will be the spectral energy distribution
measurements in these objects, with a high calibration accuracy
being the key requirement that decreases systematic errors.

\subsection{Dynamics of interstellar medium and chemical evolution of the Universe}

Chemical evolution of the Universe is of the fundamental natural
and philosophical {significance}, since the Earth and every living organism
consists of heavy elements formed in the stellar interior. Of
principal importance are both a degree of homogeneity of chemically
enriched matter [176, 177], and a homogeneity of the 
abundance {pattern as well}. For example, both from the point of view of
the conversion of the CO {line emission flux} into the column density of H$_2$
molecules, and from the origin of life in the Universe the key
question is the precise ratio of carbon to oxygen abundance, {whose spatial variations can reach an order of
magnitude in the process of chemical enrichment} depending on initial mass function of supernova
progenitors [178]. The {basic set of} observations {in framework of the study of} chemical evolution
of the Universe includes measurements of the abundance {pattern} (the
relative abundance of heavy elements), the {mass
fraction of dust}, and variations of chemical
composition {in their relations} to dynamical phenomena, such as star formation,
galactic wind, etc.

{Herschel} observations {have} demonstrated a high capability of FIR
observations to study heavy element migration in the Universe. These
observations include measurements of high-scale galactic matter
outflows,  such as galactic wind in M82 [175], observations of
bright infrared galaxies [179] and dust migration in 
galactic disks and {their outskirts} (see the discussion in
Section 6.2). Thus, {observations by Herschel} telescope provided new
insights into the nature of driving mechanisms of {such outflows},  
and {posed} several new questions about their role in the
dynamical and chemical evolution of galaxies and the Universe
{as a whole} (for a discussion see [175, 180]).

Thanks to the high angular resolution, Millimetron will be able to
measure the difference in chemical composition and 
the abundance {pattern} of interstellar gas {starting from scales where stars and supernovae inject heavy
elements out}. {Optical
observations of relative elemental abundances in the Crab nebula 
suggest that different elements show in general qualitatively
similar but quantitatively strongly distinct spatial distributions
[181].} The angular resolution of scales of the spatial variations is
relatively low (about 10-15 arcsec), therefore observations of the
Crab nebula in the IR-lines [CII] at 145 $\mu$m  and O[I] at 145
$\mu$m with an angular resolution of 3-4 arcsec are of a major
importance to estimate the degree of {chemical} inhomogeneity  at the injection scales of heavy elements into ISM.
Similar observations with the same angular resolution are possible
and interesting for other supernova remnants, for example Cas A, as
well as for nearby Wolf-Rayet stars. High angular resolution
observations of the 158 $\mu$m  and 145 $\mu$m  lines are 
essential for studies of possible spatial separation of carbon and
oxygen.

Millimetron observations in the CO lines with high principal quantum
numbers, as those of the [CII] 158 $\mu$m  and [OI] 145 $\mu$m lines
and lines of other atoms and ions, {can provide detailed investigation of 
the  enrichment of the interstellar and intergalactic medium with heavy elements.} Observations
of these lines in the direction of assumed galactic fountains in {the} 
Milky Way --- local vertical outflows driven by collective supernova
explosions in sufficiently massive OB-associations [182], are of a
high importance for studies of chemical and dynamical evolution of
the Galaxy, since they can provide {a principally}  new information on {how} heavy element are transported out of the galactic disk into
the medium and {form} their radial redistribution [183].
{Recently,} the radial distribution {of heavy elements} in the
Galaxy is has come into the focus of studies of {its} dynamical history [184]. {Among recent important results a noticeable} 
(the change by a factor three on scales from six to 12 kpc) negative
gradient of the C/O ratio [185], which can reflect peculiarities in
the chemico-dynamical evolution of the Galaxy [186], {looks worth mentioning}.

A significant fraction (half an order of magnitude) of heavy
elements is confined in solid dust particles. Therefore, 
{observations of dust} are of a principal importance not only from the point of
view of dust formation, its optical properties, {but} for studies
the chemical evolution of the Galaxy {as well}. The detection of dust in
spectra of  distant quasars at redshifts $z>6$ [187] revealed that the
dust can be formed in type II supernova shells. Details of this
process, as well as the amount of dust that can be produced by an
individual supernova, remain obscure. Recent observations of the
Crab nebula, as well as some historical type Ia supernova remnants
by the {Herschel} space telescope revealed the presence of a
significant amount of dust in these remnants, which was produced and
expelled during the supernova explosions [188, 189]. However,
reliable observational confirmations of the possibility of the dust
production by massive supernovae are absent yet (see discussion in
[190, 191]).  The possibility of observing emission from dust
particles and their seeds in the submillimeter and millimeter range
from massive supernovae in the local Universe, in particular in the
Galaxy, would be of fundamental interest.

\subsection{Gravitational lensing at high redshifts}

Observations at high redshifts of dusty star-forming galaxies (DSFGs) will be carried out to study their evolution (the redshift and luminosity determination) using homogeneous samples in the single-dish mode (photometric and spectroscopic millimeter observations).

Observations of the last decade significantly changed our
understanding of the galactic evolution by demonstrating that bright
dusty star-forming galaxies at high redhifts are about 1000 times as
numerous as in the present-day Universe (see, for example, [192]).
Spectroscopic observations of 47 sources from a high-redhsift galaxy
catalog, obtained by the South Pole Telescope (SPT) at 1.4 and 2 mm
[193], carried out  by the ALMA telescope in the frequency range
84.2-114.9 GHz showed that the brightest DSFGs are gravitationally
lensed sources [194]. The remote galaxies are lensed by foreground
galaxies in the strong gravitational lensing regime leading to
multiple images of a lensed galaxy. These results fully confirmed
the hypothesis of the gravitational lensing nature of DSFGs [195,
196].

Since the studies of properties and evolution of DSFGs are based on
observations of bright lensed sources, to determine the proper
luminosity of the source the amplification coefficient should be
known. To estimate the amplification coefficient, a full geometry of
the gravitationally lensed system should be known. To achieve this
aim, the following parameters should be measured: redshifts of the
source and the lens, relative positions of the lens end the source,
a flux ratio from the observed images, as well as galaxy-lens
properties. The latter are required to choose an adequate model of
the density distribution in the galaxy-lens.

By the launch of Millimetron, large millimeter and submillimeter sky
surveys by the Herschel telescope and SPT (220 GHz, 150 GHz) will be
completed. These surveys will be used to select lensed DSFG
candidates. Selected sample sources should meet the following main
criteria: 1) sources must have a thermal spectrum, since the dust
radiation  in the infrared and millimeter range is produced due to
re-emission of absorbed short-wavelength photons emitted by stars;
2) sources with negative K-corrections, which are the first-order
redshift corrections to a wavelength, frequency bands and
intensities, should be selected from the photometric data; 3) nearby
IR sources with $z\leq0.03$ (according to the IRAS data) and radio-loud
quasars with flat spectrum, which also emit in the millimeter range,
should be excluded.

The carbon monoxide (CO) lines are the most general indicators of
the presence of molecular gas at high redshifts. The main observable
CO transitions include $J=$ 1--0, 2--1, 3--2, 4--3, 5--4, 6--5. For example, the rest-frame emission
frequency of the transition CO $J=1-0$ is 115.27 GHz, of the transition
CO $J=5-4$ is 576.3 GHz, and of the transition CO $J=6-5$ is 691.5 GHz. The
transition CO -- produces the brightest line indicating the presence
of dense star-formation nuclei with compact morphology. For example,
from a source at $z\sim 5$, the CO $J=6-5$ transition will be observed at a
frequency of  111 GHz. For sources from such redshifts the [CII]
emission line will be shifted from 158 $\mu$m  to 948 $\mu$m .

\begin{figure}
\begin{center}
\includegraphics[width=7.5cm]{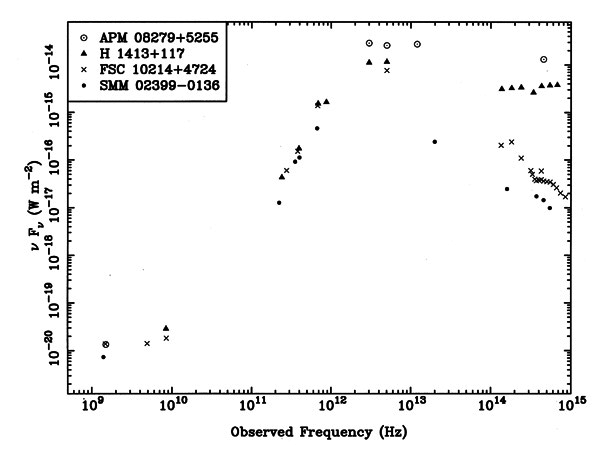}
\end{center}
\caption{Broadband spectrum of gravitational lenses [199].}
\end{figure}

Undoubtedly, the ALMA telescope will solve most of the tasks by
constructing a redshift distribution of DSFGs. However, a redhsift
range $z=1.74-2.0$ is unobservable for ALMA [193]. In addition, ALMA can
operate only within the atmosphere transparency windows, which
complicates construction of broad-band spectra. In this respect,
Millimetron has advantages in constructing broad-band millimeter and
submillimeter spectra.

In addition, it can be possible to solve the satellite problem in
the $\Lambda$CDM-model (Lambda Cold Dark Matter) by performing
photometric observations of gravitationally lensed sources with
anomalous image flux ratios  and high-resolution spectroscopic
measurements of high-redshift DSFGs in a wide frequency band.

The $\Lambda$CDM model predicts that large dark matter clumps
(halos), which host large galaxies like Milky Way should be
surrounded by several hundreds of small dark matter clumps
(subhalos) where, seemingly, dwarf satellite galaxies should be
located. However, only two-three dozen satellites have been actually
found around the Milky Way. Moreover, the Milky Way satellite
galaxies are distributed not spherically symmetric and are rather
confined within an elongated pancake tilted to the Galactic plane.

One of possible solutions of the satellite problem can be
observations of gravitationally lensed systems with anomalous image
flux ratio [197], such as MG0414+0534, MG2016+112, H1413+117 [198].
Importantly, the source H1413+117 shows a significant flux ($\sim0.1$ Jy)
at a frequency of 1 THz (Fig. 24). In addition, a pixel lensing
modeling of the gravitational lensJVAS (Jodrell/VLA Astrometric
Survey) B1938+666 in the FIR revealed the presence of satellite with
mass $10^8 M_\odot$ [200].  In this system, the source is a bright galaxy
located at $z=2.059$, the lens is a massive elliptic galaxy at $z=0.881$, and an
almost full Einstein-Khvolson ring with diameter $\sim0.9'$ is observed. The
presence of a low-mass substructure, i.e. a luminous or dark
satellite in the galaxy-lens, could locally perturb the observed
brightness distribution of the extended Einstein-Khvolson arcs.
Since these arcs are formed by multiple images of the gravitational
lens system, the ``surface brightness anomalies'' can be found and
analyzed using the pixel modeling technique and then used for the
gravitational detection, mass and location measurement of a
substructure with a mass as small as 0.1\%  of the lens mass inside
the Einstein-Khvolson ring [200].

Another way to address the ``satellite problem'' can be
high-resolution broadband spectral observations of high-redshift
DSFG lensed galaxies. Large redshifts of the observed lensed sources
enable a wide range of possible lens  redshifts to be obtained,
which in principle can constrain the evolution of subhalo population
with redshift [201]. Here of most interest are observations of the
CO molecule transition $J=6-5$, which is the brightest emission line and
indicates the presence of dense star-forming cores with compact
morphology.

A breakthrough task for the space interferometer mode can be
observations of gravitational lens candidates at high redshifts (up
to $z\sim5$) selected by submillimeter/millimeter observations in order to
confirm image multiplicity without mapping (an advantage of
Millimetron) using only the visibility function for sources with
sufficiently high brightness temperatures.

\section{Cosmology}
\subsection{Infrared background and galaxy surveys}
Galaxies at redshifts $z>1$ have a maximum flux density at
wavelengths $\lambda>200 \mu$m and produce the cosmological
IR-background (Fig. 1a). Here, the capability of Millimetron to
resolve more than 90\%  of the IR-background into individual
galaxies (Fig. 1b) can bring breakthrough results. The expected
surface density of resolved objects for photometric surveys is $\sim10^5$ per square
degree.

Massive spectral observations of a large number of galaxies will
enable to construct three-dimensional catalogues of submillimeter
galaxies at redshifts $z>2$ and to study the evolution of
large-scale dark matter distribution at an age of the Universe of
less than three billion years. Millimetron can complete 3D-catalogues
mentioned above by partial filling the gap between the recombination
era (the age of Universe 300 thousand years) and later epochs, up to
the present time (the age of Universe 3-13 billion years).

Studies of the 3D galactic space distribution, first of all, give
invaluable information on the properties of galaxies themselves: by
comparing cosmological numerical simulations with observations it is
possible to study the relation of the dark halo mass with the
observed properties of galaxies [202]. This information is important
for planning future cosmological tests from baryonic acoustic
oscillations, gravitational lensing, etc.

\subsection{Cosmological angular distances}
There are two main methods of measuring the cosmological model
parameters: geometrical and structural [203, 204]. The first method
measures the Universe expansion law, from which its geometry and
equation of state of its constituents (matter, radiation, dark
energy) can be inferred. In this method, independent measurements of
distances and redshifts of remote astronomical sources are required.

Long-term observations of water megamasers at a frequency of 22 GHz
can be used to estimate the physical sizes of accretion disks by
measuring a motion of individual maser spots. Then distances to
these objects can be derived from their angular sizes with quite a
high accuracy [205]. For the ground-based interferometers with the
maximum base this distance is limited to 300 Mpc.

Interferometry with superlong (cosmic) base can achieve an angular
resolution better than 10 micro arcsec, therefore sufficiently
bright sources can be observed from any redshift. Such distance
measurements as a function of redshift can allow to probe the
expansion law of the Universe and to study dark energy equation of
state with an unprecedentedly high accuracy.

However, this task requires multi-year very high-precision
observations, and therefore can be difficult to fulfill from the
Earth even for nearby objects. Presently, little is known about
distant megamasers to reliably judge whether is task is doable for
Millimetron.

The same relates to another similar problem. The size of a black
hole can be derived from indirect measurements of its mass. Then the
direct measurement of the black hole size can be used to determine
the cosmological distance. It is not clear as yet whether
supermassive black holes in high redshift active galactic nuclei have parameters
sufficient to be observed by Millimetron. This issue can be resolved
after observations by the EHT or RadioAstron telescopes of black
holes in the nearby galaxies and in the center of Galaxy.

\subsection{Distant galaxies and reionization of the Universe}
First stars (Population III stars) and galaxies must have been
formed from the primordial matter which had not been enriched with
heavy elements. Observations of such sources are important, firstly,
to check hypotheses of formation of first stars which reionize the
Universe, enrich the Universe by heavy chemical elements, secondly,
to understand details of the first galaxy formation and
peculiarities of the star formation in the medium with primordial
chemical composition, thirdly, to solve the supermassive black hole
formation problem.

A galaxy enriched with heavy elements
should emit in spectral lines of atomic and ionized carbon, carbon
monoxide, oxygen and other elements, and also in continuum.
Correspondingly, the primordial matter should not radiate in the
submillimeter diapason, excluding several spectral lines of simplest
molecules: HD at $112\left(1+z\right)\mu$m , $H_2$ at
$28\left(1+z\right)\mu$m , HeH${}^+$ at $149\left(1+z\right)\mu$m,
where $z$ is the galaxy redshift [15, 206, 207]. At the same time,
an atomic hydrogen emission should be observed in the near and
middle IR owing to high redshift.

Search for first galaxies is one of tasks of JWST. The search
includes the study of the dependence of the number of galaxies on
redshift $z$: the vanishing of the number of galaxies at some $z$
would indicate the galaxy formation era. However, submillimeter and
FIR observations are required to confirm the nature of such distant
objects. The lack of the detection of dust and heavy elements atoms
in the submillimeter range would suggest the discovery of a possible
primordial galaxy. The final confirmation can be obtained through
observations of the molecular lines HD (56 and 112 $\mu$m ) and H$_2$
(28, 17, 12 and 9.7 $\mu$m ), which would indicate that the gas
cooling occurs without heavy elements.

The rotational lines of molecular hydrogen, as well as of its isotopic analog HD, provide the main cooling of the energy released during gravitational contraction of the first protostellar clouds. The fluxes in these lines can much exceed the sensitivity limit of Millimetron: $\sim0.1$ mJy in the HD line at 112$(1+z)$ $\mu$m  and $\sim 1$ mJy in the H$_2$ line at 28$(1+z)$ $\mu$m , depending on the formation scenario of the first protostars. The detection of these lines is of primary importance at least to pin down the time of the star formation beginning in the Universe. In addition, measurements of these lines will help to determine or improve the redshift of first distant galaxies and quasars, which will help to construct the model of their evolution. As the simplest hydrogen molecule is well studied, it is a good indicator of physical conditions in the primordial ISM. Excitation conditions of the hydrogen molecule levels are also well studied, which enables physical conditions in the primordial gas to be probed.

The high sensitivity and availability of observations of molecular emission lines redshifted to FIR is the main advantage of Millimetron in probing the physical conditions in the primordial gas. Thus, in the near future only Millimetron will be able to study the relatively cold ($T<500$ K) ISM, including the primordial gas cooling.

\subsubsection{Spectral-space CMB perturbations}
The ``dark age'' epoch where there was no stars and galaxies ends by $z=10-25$, when first ionization sources appear: the first stars and galaxies, as well as black holes. The ionization of the Universe in this epoch can be both directly observed and inferred from the CMB polarization studies [209, 210].  However, it is unclear until now which potential ionization sources dominates. Observational studies of this problem will shed light on the formation mechanism of the first stars and supermassive black holes, whose mass can increase up to $10^9 M_\odot$ already by $z=6$, i.e. during the first billion years of the age of the Universe.

The reionization sources leads to the emerging of ionized volumes that can be probed by the kinematic Sunyaev-Zeldovich effect (the thermal Sunyaev-Zeldovich effect for these objects is much smaller). The expected level of spectral-space CMB fluctuations is $\Delta T/T=10^{-7}-10^{-6}$, which corresponds to a flux of 1-10 $\mu$Jy  in the Millimetron band. The size of the ionization region is $\sim 10$ Mpc, corresponding to an angular scale of $\sim$ 1 arcmin [211].

In addition, the emission from the HeH$^+$ molecule at redshifts $z=20-30$ falls within the CMB range, which may cause the temperature fluctuations $\Delta T/T\sim 10^{-5}$ within the spectral bands $\Delta \nu/\nu\sim 0.01$ [212].

\begin{figure}
\begin{center}
\includegraphics[width=7.5cm]{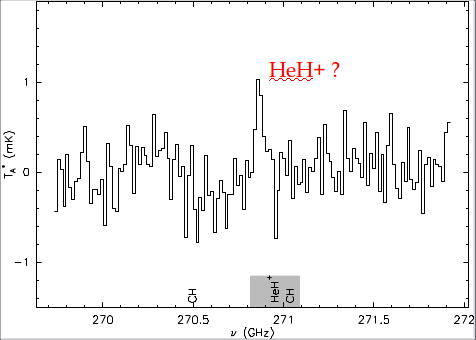}
\end{center}
\caption{Spectrum of a remote quasar at $z=6.42$ [214]. The signal-to-noise ratio $\sim 3.5$ is insufficient to reliably detect the HeH$^+$ line.}
\end{figure}

Spectral observations with low and moderate resolution in the 100-500 GHz frequency band can be used to analyze the form and evolution of the ionized clouds and to formulate constraints on the reionization scenarios: whether the Population III stars or supermassive black holes were the primary reionizaion sources.

\subsubsection{Search for emission from HeH$^+$ molecule}
Search for the HeH$^+$ molecule will help to understand details of interstellar and intergalactic medium responsible for the formation of the first sources of ionization in the early Universe.

HeH$^+$ should be one of the mostly abundant molecules in the reionization epoch [213].  This molecule is formed in the primordial gas  near the powerful ionization sources. The rest-frame wavelength of emission is 149.1 and 74.6 $\mu$m . An intriguing opportunity for the HeH$^+$ molecule searches can be provided by observations of a quasar with $z=6.42$ (the signal-to-noise ratio is 3.5) (Fig. 25).

\subsection{Galaxy clusters}
CMB photons traveling through volumes filled with sufficiently hot plasma will experience spectral changes (Fig. 26).  This is the essence of the Sunyaev-Zeldovich effect (SZ) [216, 217]. Galaxy clusters are the most appropriate for the SZ-effect observations.
Ground-based observations of the SZ-effect are restricted by the atmosphere to frequencies below 300 GHz. The use of the space telescope may lead to a breakthrough related to the separation of the thermal SZ-effect from other spectral distortions with sufficient accuracy (see Fig. 26), which can be used to measure, in particular, peculiar velocities of galaxy clusters (relative to CMB).  Measurements of these velocities will bring important information to test and improve the cosmological model [203, 204].

\begin{figure}
\begin{center}
\includegraphics[width=10.5cm]{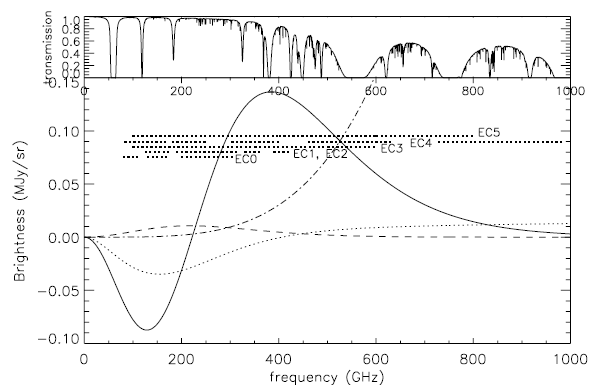}
\end{center}
\caption{(a) The Earth atmosphere transmission coefficient for dry atmosphere. (b) The Sunyaev-Zeldovich effect spectrum (deviations from the blackbody CMB emission). The solid, dashed, dotted and dash-dotted curves show the thermal effect, kinematic effect, effect on non-thermal electrons and dust emission, respectively [215].}
\end{figure}

Studies of galaxy clusters using the thermal and kinematic SZ-effect will enable the measurement of the primordial cosmological power spectrum amplitude, of the amount of dark energy in the Universe, and will help to study the growth of small perturbations and to obtain new constraints on the cosmological model, including the dark energy content. Using the SZ-effect polarization, the CMB quadrupole amplitude can be measured from the point of view of the observer co-moving with the cluster, i.e. form different regions in the Universe.

\subsection{Gamma-ray burst afterglow and host galaxies}
Despite that since the discovery of the GRB afterglow in 1997 more than 500 events have been detected in the optic and more than 800 in the X-rays, the most interesting spectral part of the afterglow remains poorly investigated. Indeed, maximum $\nu_m$ in the initial energy spectrum of the afterglow  (Fig. 27)  falls in the millimeter range. Afterglows that have been registered in this range [219] are shown in Fig. 28a. Observations of bright afterglows by Millimetron can constrain, and in some cases determine the characteristic frequency of the synchrotron emission $\nu_m$ and thus put the stringent limits on the energy, parameters of the emission region and particularly the bulk relativistic gamma-factor of an ejecta.

\begin{figure}
\begin{center}
\includegraphics[width=10.5cm]{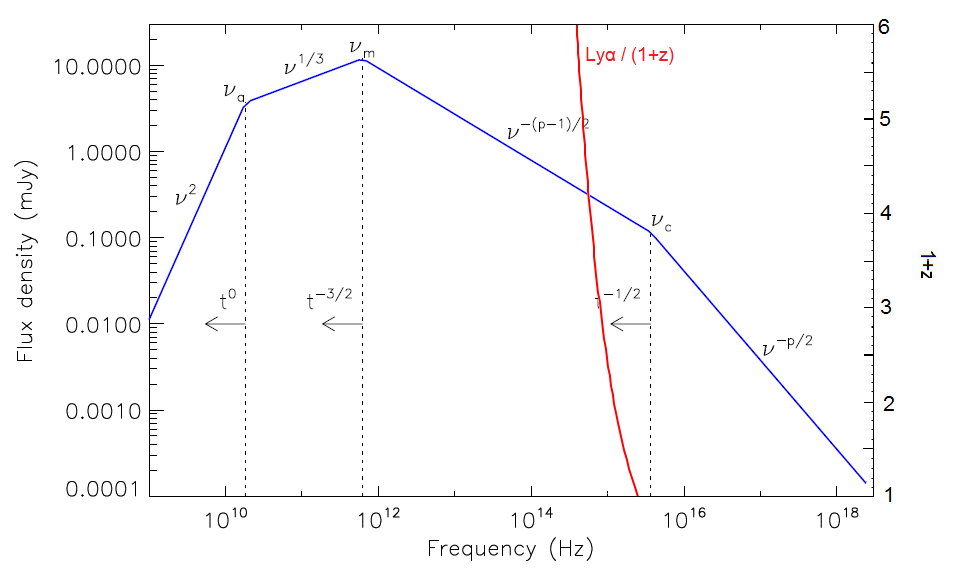}
\end{center}
\caption{Theoretical spectral energy distribution of a gamma-ray burst afterglow [218]. The red curve shows the Ly-$\alpha$ frequency in observer's frame as a function of redshift (right Y-axis). To the right of this curve, the optical emission is suppressed due to absorption in neutral hydrogen along the line of sight.}
\end{figure}

Currently, about 10\%  of the registered gamma-ray bursts have redshifts $z>5$ [222]. However, the sample of high-redshift gamma-ray bursts is incomplete because of selection effects due to difficulties to observe sources at $z\gtrsim 5$ (the maximum redshift $z=9.4$ determined for GRB 090429B [223]), where the optical emission is effectively absorbed by the Ly-$\alpha$ forest (Fig. 28a).   Therefore, the 10\%    is a lower limit of the total amount of distant GRBs.

An intriguing issue is the discrepancy between the star formation rate as inferred from distant galaxies and gamma-ray bursts (Fig. 28b).  The submillimeter observations of GRB afterglows will increase the statistics of high-redshift GRBs. On the other hand, possibly, some of already observed sources are related to explosions of the Population III stars [224, 225]. The redhsift estimate $z\geq 15$ for at least one gamma-ray burst obtained from the submillimeter observations could confirm the existence of the Population III stars and their collapses producing GRB.

Optically dark GRB, when the optic to X-ray flux ratio in the afterglow phase is extremely small are burst of special interest [226]. The redshift determination of dark gamma-ray bursts is difficult due to low (or even undetectable) optical flux. Submillimeter observations of such gamma-ray bursts   will help clarifying their nature. Indeed, the absence of the host galaxy of a gamma-ray burst in the optical (the most distant host galaxy presently known, for GRB 100219A, has $z=4.667$ [227]), and at the same time the detection of the submillimeter afterglow uniquely suggests a high redshift of the source. For close gamma-ray bursts, when redhsift can be determined by spectroscopic or photometric observations of the host galaxy, the detection of the submillimeter afterglow can be used to investigate parameters of the absorbing circumburst medium [228-231].

\begin{figure}
\begin{center}
\includegraphics[width=7.5cm]{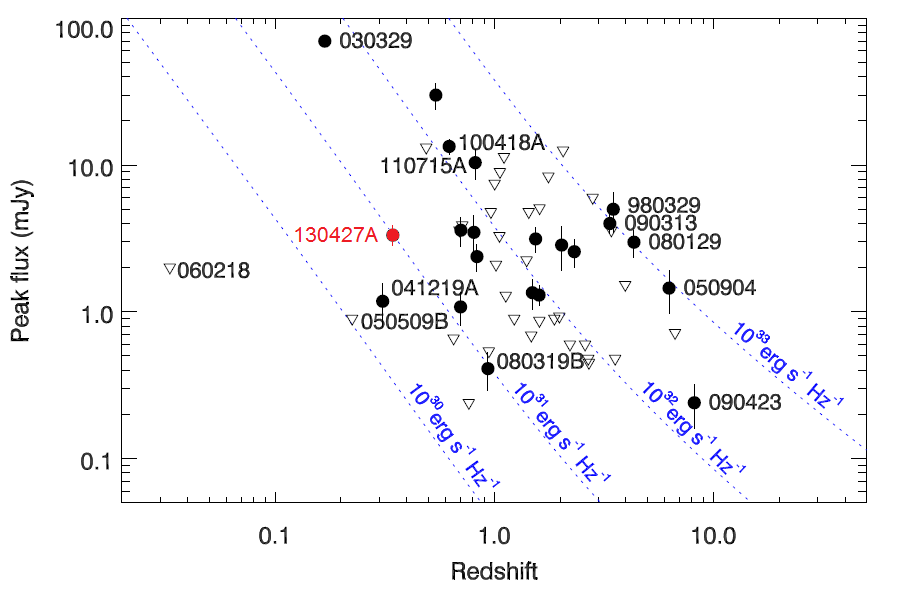}
\includegraphics[width=7.5cm]{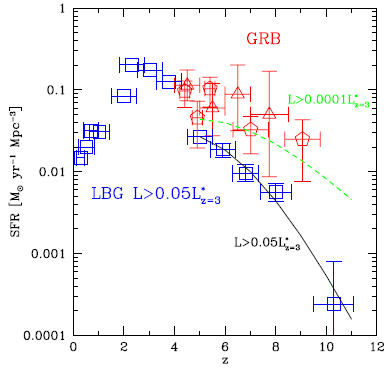}
\end{center}
\caption{(a) Submillimeter gamma-ray burst afterglows [219].  Data for GRB 130427A are taken from [220]. The dark and light symbols correspond to detection and upper limits, respectively. The dashed curves show the fluxes for sources with equal luminosity located at different redshifts. (b) The star-formation rate (SFR) as a function of redshift $z$ [221]. (GRB --- gamma-ray burst, LBG --- Lyman-Break Galaxy).}
\end{figure}

\subsection{Primordial black and white holes, wormholes and Multiverse}
The most exciting issues of the  modern cosmology include: how the Universe was born? Are there other universes and is it possible to obtain information about them? Today, this is a purely hypothetical field based on the analysis of GR equations [203, 232-238].

One of the least hypothetical assumptions is the formation of black holes in the early universe. These black holes have a non-astrophysical origin and are referred to as primordial black holes. To form them, strong density inhomogeneities at early stages of the Universe are needed [239].

The primordial black holes can have a very wide mass spectrum, and depending on mass various methods are used to their searches [240]. Millimetron will be able, as minimum, to study black holes with a mass of higher than $10^4M_\odot$  both at the present time (see Section 4.5) and in the remote past, during the reionization of the Universe (see section 7.3).

Features of the Universe passing through the singularity at the very beginning of the Big Bang can lead to the formation of interesting hypothetical objects --- wormholes, the relativistic objects similar to black holes but connecting different regions of space-time or even different universes (the Multiverse).  Possible observational distinctions between black holes and wormholes include the presence in wormholes of a radial monopole magnetic field and radiation that brings information (the image, physical parameters, etc.) from the different region of our and even from another universe [241-246]. Observational test of this hypothesis will be carried out jointly with all black hole studies (see Section 5).

General relativity also admits solutions in the form of white holes [247, 248]. Such objects can be probed, for example, by explosions like anomalous gamma-ray bursts without host galaxies [249]. Searches for host galaxies of gamma-ray bursts by Millimetron (see Section 7.5) can shed light on this possibility.

\section{Conclusions}
In conclusion, we formulate three groups of unique tasks, where
Millimetron can play a decisive role and  can strongly contribute to
solve outstanding astrophysical, astrochemical and cosmological
problems.

1. Studies of the vicinity of black holes and testing of general
relativity. Studies of accreting flows and jets near the black hole
horizons.

2. The analysis of interstellar medium and star formation. Studies
of protostars, protoplanets and protoplanetary disks, as well as of
exoplanets and the Solar system. Studies of star formation
conditions and of the interstellar medium enrichment by heavy
elements.

3. Formation end evolution of galaxies, studies of cosmological
objects and the development of the standard cosmological model.

\textbf{Acknowledgements.} The authors are deeply indebted to all whose
comments were used in the preparation of this paper: V.V. Akimkin,
A.V. Alakoz, A.A. Andrianov, N.A. Arkhipova, V.S. Beskin, M.V.
Barkov, O.V. Verkhodanov, A.A. Volnova, Yu.N. Gnedin, T. de Graauw,
V.K. Dubrovich, A.A. Ermash, D.E. Ionov, N.R. Ikhsanov, P.V.
Kaygorodov, S.V. Kalensky, A.V. Kasparova, M.S. Kirsanova, Yu.Yu.
Kovalev, S.G. Moiseenko, Y.N. Pavlyuchenko, K.A. Postnov, M.V
.Popov, O.K. Sil'chenko, V.N. Rudenko, A.V. Stepanov, S.A.
Tyulbashev, M.S. Hramtsova, N.N. Shakhvorostova, A.A. Shatsky, B.M.
Shustov, S.V. Chernov, as well as the staff of the P.N. Lebedev
Physical Institute RAS (LPI), the P.K. Sternberg State Astronomical
Institute (SAI MSU), the Institute of Astronomy RAS, the Main
Astronomical Observatory RAS, the Pushchino Radio Astronomy
Observatory and the Astro Space Center (ASC) LPI, who assisted in
the preparation of the scientific program.  The ASC LPI staff  (N.S.
Kardashev, I.D. Novikov, V.N. Lukash, S.V. Pilipenko, E.V. Mikheeva,
A.G. Doroshkevich, P.B. Ivanov, V.I. Kostenko, T.I. Larchenkova,
S.F. Likhachev, A.V. Smirnov)  thanks L.N. Likhachova for support.
A.G. Doroshkevich, P.B. Ivanov, T.I. Larchenkova, V.N. Lukash, E.V.
Miheeva, I.D. Novikov, S.V. Pilipenko acknowledge the support from
grant of the President of the Russian Federation for State Support
leading Scientific Schools NSH-4235.2014.2,  the program DPS RAS
OFN-17 "Active processes in galactic and extragalactic objects" and
the program of the Presidium of the Russian Academy of Sciences P-21
"Non-stationary phenomena in objects of the Universe". D. V.
Bisikalo and D. Z. Vibe are  supported by a grant of the President
of the Russian Federation for State Support of Leading Scientific
Schools NSH-3620.2014.2.  Yu. A. Schekinov is supported by the RFBR
grant 12-02-00917-a. The work of I.F. Malov and  V.M. Malofeev is
supported by the RFBR grant 12-02-00661 and by the Presidium of
Russian Academy of Sciences (program "The origin, structure and
evolution of objects of the Universe").  A.S. Pozanenko is supported
by the RFBR grants 12-02-01336, 13-01-92204, 14-02-10015. The work
of I.D. Novikov is supported by the RFBR grant 12-02-00276a. The
work of I.I. Zinchenko is partially supported by a grant under the
agreement between the Ministry of Education and Science of the
Russian Federation and Nizhny Novgorod State University
02.V.49.21.0003 of August 27, 2013, as well as by the RFBR grant
13-02-12220-ofi-m. A.M. Sobolev''s work was performed as part of the
job of the State Ministry of Education and Science of the Russian
Federation (project 3.1781.2014 / K). A.M. Cherepashchuk
acknowledges the grant of  the President of the Russian Federation
for State Support of Leading Scientific Schools NSH-1675.2014.2 and
RFBR grant 14-02-00825.



\begin{thebibliography}{299}
\bibitem{ar1} Dole H et al. \textit{A\&A} \textbf{451} 417 (2006)
\bibitem{ar2} Smirnov A V, Baryshev A M, Pilipenko S V, Myshonkova N V et al. \textit{Proc. SPIE 8442 Space Telescopes and Instrumentation 2012: optical, infrared and millimeter wave} eds M C Clampin et al. (2012)
\bibitem{ar3} Krabbe A, R?ser H-P \textit{Rev. of Mod. Astronomy} \textbf{12} 1052 (1999)
\bibitem{ar4} Crill B P, et al. \textit{ApJ}S \textbf{148} 527 (2003)
\bibitem{ar5} Hoogeveen, R W M, et al. \textit{Proc. of SPIE, Infrared Spaceborne Remote Sensing XI} \textbf{5152} 347 (2004)
\bibitem{ar6} Pilbratt G L, in \textit{Society of Photo-Optical Instrumentation Engineers (SPIE) Conference Series} ed. Mather J C \textbf{4850} 586 (2003)
\bibitem{ar7} Kardashev N S et al. Trudy FIAN 228 112 (2000)
\bibitem{ar8} Wild W, Kardashev N S et al. \textit{Exp Astron} \textbf{23} 221 (2009)
\bibitem{ar9} Kiuchi H Coherence estimation on the measured phase noise in Allan standard deviation ALMA Memo \textbf{530} (2005)
\bibitem{ar10} Nakagawa T, in \textit{Society of Photo-Optical Instrumentation Engineers (SPIE) Conference Series} 7731 (2010)
\bibitem{ar11} Doeleman S S et al. \textit{Nature} \textbf{455} 78 (2008)
\bibitem{ar12} McKee and Ostriker \textit{ARAA} \textbf{45} 565 (2007)
\bibitem{ar13} Goicoechea J R and Cernicharo J \textit{ApJ} \textbf{554} L213 (2001)
\bibitem{ar14} Planck Collaboration \textit{A\&A} \textbf{536} 19 (2011)
\bibitem{ar15} Roberge W, Dalgarno A \textit{ApJ} \textbf{255} 489 (1982)
\bibitem{ar16} Cecchi-Pestellini C., Dalgarno A., \textit{ApJ} \textbf{413} 611 (1993)
\bibitem{ar17} Voshchinnikov N V, private communication
\bibitem{ar18} Efremov Yu N \textit{MNRAS} \textbf{405} 1531 (2010)
\bibitem{ar19} Fukui Y, Kawamura A \textit{ARA\&A} \textbf{48} 547 (2010)
\bibitem{ar20} Fukui Y et al. \textit{ApJ} \textbf{780} 36 (2014)
\bibitem{ar21} Inutsuka S-I, Koyama H \textit{Ap\&SS} \textbf{281} 67
\bibitem{ar22} Khoperskov S A, Vasiliev E O, Sobolev A M and Khoperskov A V \textit{MNRAS} \textbf{428} 2311 (2013)
\bibitem{ar23} Pineda J L, Langer W D, Velusamy T, и Goldsmith P F \textit{A\&A} \textbf{554} 103 (2013)
\bibitem{ar24} Kirsanova M S, Sobolev A M, Thomasson M, Wiebe D S, Johansson L E B, Seleznev A F \textit{MNRAS} \textbf{388} 729 (2008)
\bibitem{ar25} Kirsanova M S, Wiebe D S, Sobolev A M, Henkel C, Tsivilev A P \textit{MNRAS} \textbf{437} 1593 (2014)
\bibitem{ar26} Crockett N R et al. \textit{A\&A} \textbf{521} L21 (2010)
\bibitem{ar27} http://www.sron.rug.nl/millimetron/OxygenPuzzle
\bibitem{ar28} Walmsley M, van der Tak F in procs. ``Dusty and molecular universe: a prelude to Herschel and ALMA'' ESA \textbf{SP-577} 55 (2005)
\bibitem{ar29} Wyrowski F. et al. \textit{A\&A} \textbf{542} L15 (2012)
\bibitem{ar30} Kirsanova M S, Wiebe D S, Sobolev A M Astron. Rep. \textbf{53} 611 (2009)
\bibitem{ar31} Bergin E A et al. \textit{Nature} \textbf{493} 644 (2013)
\bibitem{ar32} Bergin E A, Hogerheijde M R, Brinch C, Fogel J et al. \textit{Astron. Astrophys} \textbf{521} L33 (2010)
\bibitem{ar33} Hogerheijde M R, Bergin E A, Brinch Ch, Cleeves L I \textit{Science} 334 338 (2011)
\bibitem{ar34} Lahuis F, van Dishoeck E F, Boogert A C A, Pontoppidan K M \textit{Astrophys J} \textbf{636} L145 (2006)
\bibitem{ar35} Akimkin V et al. \textit{ApJ} \textbf{766} 8 (2013)
\bibitem{ar36} Akimkin V et al. \textit{Astrophys. Space Sci.} \textbf{335} 33 (2011)
\bibitem{ar37} Young Ch H et al. \textit{Astrophys. J. Suppl.} \textbf{154} 396 (2004)
\bibitem{ar38} Pavlyuchenkov Ya N, Vibe D S, Fateeva A M, Vsyunina T S \textit{Astron. Zurn.} \textbf{88} 3 (2011)
\bibitem{ar39} Robitaille Th O et al. \textit{Astrophys. J. Suppl.} \textbf{167} 256 (2006)
\bibitem{ar40} Robitaille Th P et al. \textit{Astrophys. J. Suppl.} \textbf{169} 328 (2007)
\bibitem{ar41} Zinchenko I I, in preparation (2014)
\bibitem{ar42} Henning Th et al. \textit{Astron. Astrophys.} \textbf{518} L95 (2010)
\bibitem{ar43} Muller T G et al. \textit{Astron. Astrophys.} \textbf{566} A22 (2014); arXiv:1404.5847
\bibitem{ar44} Gray M Maser Sources in Astrophysics (Cambridge: Cambridge Univ. Press, 2012)
\bibitem{ar45} Sobolev AM et al. \textit{Proc. Int. Astron. Union} \textbf{242} 81 (2007)
\bibitem{ar46} Moran J Met al. \textit{Proc. Int. Astron. Union} \textbf{242} 391 (2007)
\bibitem{ar47} Parfenov S Yu, Sobolev A M \textit{Mon. Not. R. Astron. Soc.} \textbf{444} 620 (2014)
\bibitem{ar48} Zhilkin AG, BisikaloDV, Boyarchuk A A \textit{Phys. Usp.} \textbf{55} 115 (2012)
\bibitem{ar49} Matveyenko L I et al. \textit{Astron. Lett.} \textbf{30} 100 (2004)
\bibitem{ar50} Sobolev AM et al. \textit{Science} (2015), in print
\bibitem{ar51} Bisikalo D V \textit{Phys. Usp.} \textbf{51} 551 (2008)
\bibitem{ar52} Kurbatov E P, Bisikalo D V, Kaygorodov P V \textit{Phys. Usp.} \textbf{57} 787 (2014)
\bibitem{ar53} Seager S Exoplanet Atmospheres: Physical Processes (Princeton, N.J.: Princeton Univ. Press, 2010)
\bibitem{ar54} Vidal-Madjar A et al. \textit{Nature} \textbf{422} 143 (2003)
\bibitem{ar55} Linsky J et al. \textit{Astrophys. J.} \textbf{717} 1291 (2010)
\bibitem{ar56} France K et al. \textit{Astrophys. J.} \textbf{712} 1277 (2010)
\bibitem{ar57} Fossati L et al. \textit{Astrophys. J.} \textbf{714} L222 (2010)
\bibitem{ar58} Narita N et al. \textit{Astrophys. J.} \textbf{773} 144 (2013)
\bibitem{ar59} Mayer A et al. \textit{Astron. Astrophys.} \textbf{549} A69 (2013)
\bibitem{ar60} Decin L \textit{Adv. Space Res.} \textbf{50} 843 (2012)
\bibitem{ar61} Justtanont K et al. \textit{Astron. Astrophys.} \textbf{537} 144 (2012)
\bibitem{ar62} Cernicharo J et al. \textit{ApJ} \textbf{778} L25 (2013)
\bibitem{ar63} Ueta T et al. \textit{Astron. Astrophys.} \textbf{565} A36 (2014)
\bibitem{ar64} Kardashev N S \textit{Nature} \textbf{278} 28 (1979)
\bibitem{ar65} Mauersberger R et al. \textit{Astron. Astrophys.} \textbf{306} 141 (1996)
\bibitem{ar66} Dyson F \textit{Science} \textbf{131} 1667 (1960)
\bibitem{ar67} Slysh V I, in The Search for Extraterrestrial Life --- Recent Developments: Proc. of the 112th Symp. of the Intern. Astronomical Union, Boston, Mass., USA, June 18 -- 21, 1984 (Ed M D Papagiannis)(Dordrecht: D. Reidel, 1985) p. 315
\bibitem{ar68} Carrigan R \textit{Astrophys. J.} \textbf{698} 2075 (2009)
\bibitem{ar69} Kilic Met al. \textit{Astrophys. J.} \textbf{678} 1298 (2008)
\bibitem{ar70} Farihi J et al. \textit{Mon. Not. R. Astron. Soc.} \textbf{432} 1955 (2013)
\bibitem{ar71} CutriRMet al. ``WISE All-sky Data Release'', 2012yCat.2311....0C (2012)
\bibitem{ar72} Kawka A, Vennes S \textit{Mon. Not. R. Astron. Soc. Lett.} \textbf{439} L90 (2014)
\bibitem{ar73} Malofeev V et al. \textit{Astron. Astrophys.} \textbf{285} 201 (1994)
\bibitem{ar74} Xilouris KMet al. \textit{Astron. Astrophys.} \textbf{288} L17 (1994)
\bibitem{ar75} Morris D et al. \textit{Astron. Astrophys.} \textbf{322} L17 (1997)
\bibitem{ar76} Lohmer O et al. \textit{Astron. Astrophys.} \textbf{480} 623 (2008)
\bibitem{ar77} Kramer Met al. \textit{Astrophys. J.} \textbf{488} 364 (1997)
\bibitem{ar78} Camilo F et al. \textit{Astrophys. J.} \textbf{669} 561 (2007)
\bibitem{ar79} Camilo F et al. \textit{Astrophys. J.} \textbf{679} 681 (2008)
\bibitem{ar80} Malov I F \textit{Astron. Rep.} \textbf{41} 617 (1997)
\bibitem{ar81} Malov I F \textit{Astron. Rep.} \textbf{58} 139 (2014)
\bibitem{ar82} Popov M V et al. \textit{Astrophys. J. Lett.} (2015), in print
\bibitem{ar83} Heger A et al. \textit{Astrophys. J.} \textbf{591} 288 (2003)
\bibitem{ar84} Marscher A P et al. \textit{Astrophys. J.} \textbf{763} L15 (2013)
\bibitem{ar85} Mackey A D et al. \textit{Mon. Not. R. Astron. Soc.} \textbf{386} 65 (2008)
\bibitem{ar86} Strader J et al. \textit{Nature} \textbf{490} 71 (2012)
\bibitem{ar87} Chomiuk L et al. \textit{Astrophys. J.} \textbf{777} 69 (2013); arXiv:1306.6624
\bibitem{ar88} Fabbiano G \textit{Annu. Rev. Astron. Astrophys.} \textbf{27} 87 (1989)
\bibitem{ar89} Frank J, King A, Raine D J Accretion Power in Astrophysics (Cambridge: Cambridge Univ. Press, 2002)
\bibitem{ar90} Colbert E J M, Mushotzky R F \textit{Astrophys. J.} \textbf{519} 89 (1999)
\bibitem{ar91} Liu J-F, Bregman J N \textit{Astrophys. J. Suppl.} \textbf{157} 59 (2005)
\bibitem{ar92} Liu Q Z, Mirabel I F \textit{Astron. Astrophys.} \textbf{429} 1125 (2005)
\bibitem{ar93} Colbert E J M, Ptak A F \textit{Astrophys. J. Suppl.} \textbf{143} 25 (2002)
\bibitem{ar94} Swartz D A \textit{ASP Conf. Ser.} \textbf{423} 277 (2010)
\bibitem{ar95} Mushotzky R \textit{Prog. Theor. Phys. Suppl.} \textbf{155} 27 (2004)
\bibitem{ar96} Miller JMet al. \textit{Astrophys. J. Lett.} \textbf{585} L37 (2003)
\bibitem{ar97} Poutanen J et al. \textit{Mon. Not. R. Astron. Soc.} \textbf{377} 1187 (2007)
\bibitem{ar98} Kording E, Falcke H, Markoff S \textit{Astron. Astrophys.} \textbf{382} L13 (2002)
\bibitem{ar99} Feng H, Soria R \textit{New Astron. Rev.} \textbf{55} 166 (2011)
\bibitem{ar100} Miller M, Coleman H, Douglas P \textit{Astrophys. J.} \textbf{576} 894 (2002)
\bibitem{ar101} Farrell S A et al. \textit{Nature} \textbf{460} 73 (2009)
\bibitem{ar102} Webb N et al. \textit{Science} \textbf{337} 554 (2012)
\bibitem{ar103} MillerNA, Mushotzky R F, Neff SG \textit{Astrophys. J.} \textbf{623} L109 (2005)
\bibitem{ar104} Kaaret P et al. \textit{Science} \textbf{299} 365 (2003)
\bibitem{ar105} Harrison F, in Intern. Conf. ZELDOVICH-100, Moscow, Russia, 16 -- 20 June 2014
\bibitem{ar106} Bisnovatyi-Kogan G S \textit{Sov. Astron.} \textbf{14} 652 (1971)
\bibitem{ar107} Moiseenko S G, Bisnovatyi-KoganGS, ArdeljanN V \textit{Mon. Not. R. Astron. Soc.} \textbf{370} 501 (2006)
\bibitem{ar108} Bisnovatyi-Kogan G S, Moiseenko S G, Ardelyan N V \textit{Astron. Rep.} \textbf{52} 997 (2008)
\bibitem{ar109} Bisnovatyi-Kogan G S, Moiseenko S G, Ardelyan N V \textit{Astron. Rep.} \textbf{36} 285 (1992)
\bibitem{ar110} Barkov M V, Komissarov S S \textit{Mon. Not. R. Astron. Soc.} \textbf{415} 944 (2011)
\bibitem{ar111} Chevalier R A \textit{Astrophys. J. Lett.} \textbf{752} 2 (2012)
\bibitem{ar112} Taam R E, Sandquist E L \textit{Astron. Astrophys.} \textbf{38} 113 (2000)
\bibitem{ar113} Field G B, Rogers R D \textit{Astrophys. J.} \textbf{403} 94 (1993)
\bibitem{ar114} Kardashev N S \textit{Mon. Not. R. Astron. Soc.} \textbf{276} 515 (1995)
\bibitem{ar115} Lobanov A P \textit{Astron. Astrophys.} \textbf{330} 79 (1998)
\bibitem{ar116} Tyul'bashev S A \textit{Astron. Astrophys.} \textbf{387} 818 (2002)
\bibitem{ar117} Eatough et al. \textit{Nature} \textbf{501} 391 (2013)
\bibitem{ar118} Marrone et al. \textit{Astrophys. J. Lett.} \textbf{654} 57 (2007)
\bibitem{ar119} Trippe et al. \textit{Astron. Astrophys.} \textbf{540} A74 (2012)
\bibitem{ar120} Ivanov P B, private communication
\bibitem{ar121} Dexter J et al. \textit{Astrophys. J.} \textbf{717} 1092 (2010)
\bibitem{ar122} Lu Y et al. \textit{Astrophys. J.} (2014), in print
\bibitem{ar123} Muller T, Frauendiener J \textit{Eur. J. Phys.} \textbf{33} 955 (2012)
\bibitem{ar124} Komossa S, Bade N \textit{Astron. Astrophys.} \textbf{343} 775 (1999)
\bibitem{ar125} Bloom J S et al. \textit{GCN Circ.} (11847) 1 (2011)
\bibitem{ar126} Beskin V S \textit{Phys. Usp.} \textbf{53} 1199 (2010)
\bibitem{ar127} Barkov M V, Komissarov S S \textit{Mon. Not. R. Astron. Soc.} \textbf{401} 1644 (2010)
\bibitem{ar128} Barkov MV, Baushev A N \textit{New Astron.} \textbf{16} 45 (2011)
\bibitem{ar129} McKinney J C, Uzdensky D A \textit{Mon. Not. R. Astron. Soc.} \textbf{419} 573 (2012)
\bibitem{ar130} McKinney J C, Tchekhovskoy A, Blandford R D \textit{Mon. Not. R. Astron. Soc.} \textbf{423} 3083 (2012)
\bibitem{ar131} De Bruyn A G NFRA Note 655 1 (1996)
\bibitem{ar132} BrentjensMA, de BruynAG, in Proc. of the Riddle of Cooling Flows in Galaxies and Clusters of Galaxies, Charlottesville, VA, USA, May 31 -- June 4, 2003 (Eds T H Reiprich, J C Kempner, N Soker) (Charlottesville, VA: Univ. of Virginia, 2004)
\bibitem{ar133} Vogt C, Dolag K, Enslin T A \textit{Mon. Not. R. Astron. Soc.} \textbf{358} 726 (2005)
\bibitem{ar134} Brentjens MA, de Bruyn A G \textit{Astron. Astrophys.} \textbf{441} 1217 (2005)
\bibitem{ar135} Aharonian F et al. \textit{Astrophys. J.} \textbf{664} L71 (2007)
\bibitem{ar136} Begelman M C, Fabian A C, Rees M J \textit{Mon. Not. R. Astron. Soc.} \textbf{384} L19 (2008)
\bibitem{ar137} Giannios D, Uzdensky D A, Begelman M C \textit{Mon. Not. R. Astron. Soc.} \textbf{395} L29 (2009)
\bibitem{ar138} Barkov MV et al. \textit{Mon. Not. R. Astron. Soc.} \textbf{749} 119 (2012)
\bibitem{ar139} Mesler R A, Pihlstrom YM \textit{Astrophys. J.} \textbf{774} 77 (2013)
\bibitem{ar140} Berger E et al. \textit{Nature} \textbf{426} 154 (2003)
\bibitem{ar141} Frail D A et al. \textit{Astrophys. J.} \textbf{619} 994 (2005)
\bibitem{ar142} Barkov M V, Pozanenko A S \textit{Mon. Not. R. Astron. Soc.} \textbf{417} 2161 (2011)
\bibitem{ar143} Michalowski M, Hjorth J, Watson D \textit{Astron. Astrophys.} \textbf{514} A67 (2010)
\bibitem{ar144} Pope A, Chary R R \textit{Astrophys. J.} \textbf{715} L171 (2010)
\bibitem{ar145} Dayal P, Hirashita H, Ferrara A \textit{Mon. Not. R. Astron. Soc.} \textbf{403} 620 (2010)
\bibitem{ar146} Pohlen Met al. \textit{Astron. Astrophys.} \textbf{518} L72 (2010)
\bibitem{ar147} Fritz Met al. \textit{Astron. Astrophys.} \textbf{516} 34 (2012)
\bibitem{ar148} Smith MWL et al. \textit{Astrophys. J.} \textbf{756} 40 (2012)
\bibitem{ar149} Holwerda BWet al. \textit{Astron. Astrophys.} \textbf{444} 101 (2005)
\bibitem{ar150} Holwerda BWet al. \textit{Astron. Nachr.} \textbf{334} 268 (2013)
\bibitem{ar151} Magrini L et al. \textit{Astron. Astrophys.} \textbf{535} 13 (2011)
\bibitem{ar152} Grenier I A, Casandjian J M, Terrier R \textit{Science} \textbf{307} 1292 (2005)
\bibitem{ar153} Abramova O V, Zasov A V \textit{Astron. Lett.} \textbf{38} 755 (2012)
\bibitem{ar154} Karachentsev I et al. \textit{Mon. Not. R. Astron. Soc.} \textbf{415} L31 (2011)
\bibitem{ar155} Das M, Boone F, Viallefond F \textit{Astron. Astrophys.} \textbf{523} 63 (2010)
\bibitem{ar156} Hinz J L et al. \textit{Astrophys. J.} \textbf{663} 895 (2007)
\bibitem{ar157} Kasparova A V et al. \textit{Mon. Not. R. Astron. Soc.} \textbf{437} 3072 (2014)
\bibitem{ar158} Xilouris E et al. \textit{Astrophys. J.} \textbf{651} L107 (2006)
\bibitem{ar159} Temi P, Brighenti F, Mathews WG \textit{Astrophys. J.} \textbf{660} 1215 (2007)
\bibitem{ar160} Agguirre A, Haiman Z \textit{Astrophys. J.} \textbf{532} 28 (2000)
\bibitem{ar161} Johansson J, Morstell E \textit{Mon. Not. R. Astron. Soc.} \textbf{426} 3360 (2012)
\bibitem{ar162} Nath B, Shchekinov Yu \textit{Rep. Prog. Phys.} (2014), submitted
\bibitem{ar163} Polikarpova O L, Shchekinov Yu A \textit{Astron. Rep.} In press (2014)
\bibitem{ar164} Chelouche D, Koester B P, BowenDV \textit{Astrophys. J.} \textbf{671} L97 (2007)
\bibitem{ar165} Muller S et al. \textit{Astrophys. J.} \textbf{680} 975 (2008)
\bibitem{ar166} Menard B, Kilbinger M, Scranton R \textit{Mon. Not. R. Astron. Soc.} \textbf{406} 1815 (2010)
\bibitem{ar167} McGee S L, BaloghML \textit{Mon. Not. R. Astron. Soc.} \textbf{405} 2069 (2010)
\bibitem{ar168} Kitayama T et al. \textit{Astrophys. J.} \textbf{695} 1191 (2009)
\bibitem{ar169} Grossi Met al. \textit{Astron. Astrophys.} \textbf{518} L52 (2010)
\bibitem{ar170} BaesM et al. \textit{Astron. Astrophys.} \textbf{518} L53 (2010)
\bibitem{ar171} Simcoe R A et al. \textit{Astrophys. J.} \textbf{637} 648 (2006)
\bibitem{ar172} Tumlinson J et al. \textit{Science} \textbf{334} 948 (2011)
\bibitem{ar173} Vasiliev E O, Nath B B, Shchekinov Yu \textit{Mon. Not. R. Astron. Soc.} (2014), submitted; arXiv:1401.5070
\bibitem{ar174} Khramtsova MS et al. \textit{Mon. Not. R. Astron. Soc.} \textbf{431} 2006 (2013)
\bibitem{ar175} Contursi A et al. \textit{Astron. Astrophys.} \textbf{549} 118 (2013)
\bibitem{ar176} Dedikov S Yu, Shchekinov Yu A \textit{Astron. Rep.} \textbf{48} 9 (2004)
\bibitem{ar177} Vasiliev E O, Dedikov S Yu, Shchekinov Yu A \textit{Astrophys. Bull.} \textbf{64} 317 (2009)
\bibitem{ar178} Shchekinov Yu A, Vasiliev E O \textit{Astrophys. Space Sci.} (2014), submitted
\bibitem{ar179} Diaz-Santos T et al. \textit{Astrophys. J. Lett.} \textbf{788} L17 (2014)
\bibitem{ar180} Kreckel K et al. \textit{Astrophys. J.} \textbf{790} 26 (2014)
\bibitem{ar181} Satteréeld T J et al. \textit{Astron. J.} \textbf{144} 27 (2012)
\bibitem{ar182} Heiles C, ReachW T, Koo B-C \textit{Astrophys. J.} \textbf{466} 191 (1996)
\bibitem{ar183} Tenorio-Tagle G \textit{Astron. J.} \textbf{111} 1641 (1996)
\bibitem{ar184} Lemasle B et al. \textit{Astron. Astrophys.} \textbf{558} 31 (2013)
\bibitem{ar185} Esteban C et al. \textit{Mon. Not. R. Astron. Soc.} \textbf{433} 382 (2013)
\bibitem{ar186} Esteban C et al. \textit{Mon. Not. R. Astron. Soc.} \textbf{443} 624 (2014)
\bibitem{ar187} Maiolino R et al. \textit{Nature} \textbf{431} 533 (2004)
\bibitem{ar188} Gomez H L et al. \textit{Astrophys. J.} \textbf{760} 96 (2012)
\bibitem{ar189} Gomez H L et al. \textit{Mon. Not. R. Astron. Soc.} \textbf{420} 3557 (2012)
\bibitem{ar190} Krause O et al. \textit{Nature} \textbf{432} 596 (2004)
\bibitem{ar191} Barlow MJ et al. \textit{Astron. Astrophys.} \textbf{518} L138 (2010)
\bibitem{ar192} Lagache G, Puget J-L, Dole H Annu. \textit{Rev. Astron. Astrophys.} \textbf{43} 727 (2005)
\bibitem{ar193} Vieira J D et al. \textit{Astrophys. J.} \textbf{719} 763 (2010)
\bibitem{ar194} Vieira J D et al. \textit{Nature} \textbf{495} 344 (2013); arXiv:1303.2723
\bibitem{ar195} Blain A W \textit{Mon. Not. R. Astron. Soc.} \textbf{283} 1340 (1996)
\bibitem{ar196} Negrello Met al. \textit{Mon. Not. R. Astron. Soc.} \textbf{377} 1557 (2007)
\bibitem{ar197} Mao S D, Kochanek C S \textit{Mon. Not. R. Astron. Soc.} \textbf{268} 569 (1994)
\bibitem{ar198} McLeod K K, Bechtold J \textit{Astrophys. J.} \textbf{704} 415 (2009)
\bibitem{ar199} Irwin MJ et al. \textit{Astrophys. J.} \textbf{505} 529 (1998)
\bibitem{ar200} Vegetti S et al. \textit{Nature} \textbf{481} 341 (2012)
\bibitem{ar201} Hezaveh Y D et al. \textit{Astrophys. J.} \textbf{761} 20 (2012)
\bibitem{ar202} Viero MP et al. \textit{Astrophys. J.} \textbf{772} 77 (2013)
\bibitem{ar203} Lukash V N, Mikheeva E V Physical cosmology (M.: Fizmatlit, 2010)
\bibitem{ar204} Lukash V N, Mikheeva E V, Malinovsky A M \textit{Phys. Usp.} \textbf{54} 983 (2011)
\bibitem{ar205} Braatz R et al. \textit{Astrophys. J.} \textbf{767} 154 (2013)
\bibitem{ar206} Galli D, Palla F \textit{Annu. Rev. Astron. Astrophys.} \textbf{51} 163 (2013)
\bibitem{ar207} Shchekinov Yu A \textit{Sov. Astron. Lett.} \textbf{12} 211 (1986)
\bibitem{ar208} Kamaya H, Silk J \textit{Mon. Not. R. Astron. Soc.} \textbf{332} 251 (2009)
\bibitem{ar209} Becker R H et al. \textit{Astron. J.} \textbf{122} 2850 (2001)
\bibitem{ar210} Dunkley J et al. \textit{Astrophys. J. Suppl.} \textbf{180} 306 (2009)
\bibitem{ar211} Doroshkevich A G, Pilipenko S V \textit{Astron. Rep.} \textbf{55} 567 (2011)
\bibitem{ar212} Dubrovich V K, Bajkova A, Khaikin V B \textit{New Astron.} \textbf{13} 28 (2008)
\bibitem{ar213} Dubrovich V K, Lipovka A A \textit{Astron. Astrophys.} \textbf{296} 307 (1995)
\bibitem{ar214} Zinchenko I, Dubrovich V, Henkel C \textit{Mon. Not. R. Astron. Soc.} \textbf{415} L78 (2011)
\bibitem{ar215} de Bernardis P et al. \textit{Astron. Astrophys.} \textbf{538} 86 (2012)
\bibitem{ar216} Zeldovich Ya B, Sunyaev R A \textit{Astrophys. Space Sci.} \textbf{288} 4 (1969)
\bibitem{ar217} Zeldovich Ya B, Sunyaev R A \textit{Astrophys. Space Sci.} \textbf{20} 9 (1969)
\bibitem{ar218} Sari R, Piran T, Narayan R \textit{Astrophys. J.} \textbf{497} L17 (1998)
\bibitem{ar219} De Ugarte Postigo et al. \textit{Astron. Astrophys.} \textbf{538} 44 (2012)
\bibitem{ar220} Perley D A et al. \textit{Astrophys. J.} \textbf{781} 37 (2014)
\bibitem{ar221} Trenti Met al. \textit{Astrophys. J. Lett.} \textbf{749} L38 (2012)
\bibitem{ar222} Tanvir N R et al. \textit{Astrophys. J.} \textbf{754} 46 (2012)
\bibitem{ar223} Cucchiara A et al. \textit{Astrophys. J.} \textbf{736} 7 (2011)
\bibitem{ar224} Bromm V, Coppi P S, Larson R B \textit{Astron. J.} \textbf{564} 23 (2002)
\bibitem{ar225} Komissarov S S, Barkov M V \textit{Mon. Not. R. Astron. Soc.} \textbf{402} L25 (2010)
\bibitem{ar226} Jakobsson P et al. \textit{Astrophys. J.} \textbf{617} L21 (2004)
\bibitem{ar227} Thoene C C et al. \textit{Mon. Not. R. Astron. Soc.} \textbf{428} 3590 (2013)
\bibitem{ar228} Castro-Tirado A J et al. \textit{Astron. Astrophys.} \textbf{475} 101 (2007)
\bibitem{ar229} Perley D A et al. \textit{Astron. J.} \textbf{138} 1690 (2009)
\bibitem{ar230} Zauderer B A et al. \textit{Astrophys. J.} \textbf{767} 161 (2013)
\bibitem{ar231} Volnova A A et al. \textit{Mon. Not. R. Astron. Soc.} \textbf{442} 2586 (2014)
\bibitem{ar232} Linde A Particle Physics and Inflationary Cosmology (Chur: Harwood Acad. Publ., 1990)
\bibitem{ar233} Linde A, in Universe or Multiverse? (Ed. B Carr) (Cambridge: Cambridge Univ. Press, 2007) p. 127
\bibitem{ar234} Mukhanov V F Physical Foundations of Cosmology (Cambridge: Cambridge Univ. Press, 2007)
\bibitem{ar235} Smolin L The Life of the Cosmos (Oxford: Oxford Univ. Press, 1999)
\bibitem{ar236} Lukash V N, Mikheeva E V, Strokov V N \textit{Phys. Usp.} \textbf{55} 831 (2012)
\bibitem{ar237} Lukash V N, Mikheeva E V, Strokov V N \textit{Phys. Usp.} \textbf{55} 204 (2012)
\bibitem{ar238} Garay I, Robles-Perez S \textit{Int. J. Mod. Phys. D} \textbf{23} 1450043 (2014)
\bibitem{ar239} Nadezhin D K, Novikov I D, Polnarev A G \textit{Sov. Astron.} \textbf{22} 129 (1978)
\bibitem{ar240} Carr B J et al. \textit{Phys. Rev. D} \textbf{81} 104019 (2010)
\bibitem{ar241} Kardashev N S, Novikov I D, Shatskii A A \textit{Astron. Rep.} \textbf{50} 601 (2006)
\bibitem{ar242} Doroshkevich A G, Kardashev N S, Novikov D I, Novikov I D \textit{Astron. Rep.} \textbf{52} 616 (2008)
\bibitem{ar243} Pozanenko A, Shatskiy A \textit{Gravit. Cosmol.} \textbf{16} 259 (2010); arXiv:1007.3620
\bibitem{ar244} Shatskii A A, Novikov I D, Kardashev N S \textit{Phys. Usp.} \textbf{51} 457 (2008)
\bibitem{ar245} Novikov I D, KardashevNS, Shatskii A A \textit{Phys. Usp.} \textbf{50} 965 (2007)
\bibitem{ar246} Shatskii A A \textit{Phys. Usp.} \textbf{52} 811 (2009)
\bibitem{ar247} Novikov I D \textit{Sov. Astron.} \textbf{8} 857 (1965)
\bibitem{ar248} Novikov I D, Frolov V P Physics of Black Holes (Dordrecht: Kluwer Acad., 1989)
\bibitem{ar249} Retter A, Heller S \textit{New Astron.} \textbf{17} 73 (2012)
\end{thebibliography}
\end{document}